\documentclass{emulateapj} 
\usepackage{graphicx}
\usepackage{apjfonts}
\usepackage{ifthen}

\def\lsim{\mathrel{\rlap{\lower4pt\hbox{\hskip1pt$\sim$}}
    \raise1pt\hbox{$<$}}}                
\def\gsim{\mathrel{\rlap{\lower4pt\hbox{\hskip1pt$\sim$}}
    \raise1pt\hbox{$>$}}}                

\def\xt{\times 10^}

\usepackage{ulem}
\usepackage{color}
\newif\ifTRACKCHANGES
\TRACKCHANGESfalse
\ifTRACKCHANGES
   \newcommand{ \removethis}[ 1]{ \textcolor{red}{ #1 }}
   \newcommand{ \addthis}[ 1]{ \textcolor{blue}{ #1 }}
\else
   \newcommand{ \removethis}[ 1]{}
   \newcommand{ \addthis}[ 1]{#1}
\fi

\newboolean{colver}
\setboolean{colver}{true}

\slugcomment{Received 2009, September 22; accepted 2010, January 11.}

\begin{document}


\title{A Larger Estimate of the Entropy of the Universe}

\author{Chas A.\ Egan}
\affil{Research School of Astronomy and Astrophysics, Australian National University, Canberra, Australia \altaffilmark{1}} 
\altaffiltext{1}{Department of Astrophysics, School of Physics, University of New 
South Wales, Sydney, Australia}
\email{chas@mso.anu.edu.au}

\author{Charles H.\ Lineweaver}
\affil{Planetary Science Institute, Research School of Astronomy and Astrophysics and Research School of Earth Sciences, 
Australian National University, Canberra, Australia} 




\begin{abstract}
	Using recent measurements of the supermassive black hole (SMBH) mass function, we find that SMBHs are 
	the largest contributor to the entropy of the observable universe, contributing at least
	an order of magnitude more entropy than previously estimated. The total entropy 
	of the observable universe is correspondingly higher, and is $S_{\mathrm{obs}} = 3.1^{+3.0}_{-1.7}\xt{104} k$.
	We calculate the entropy of the current cosmic event horizon to be 
	$S_{\mathrm{CEH}} = 2.6 \pm 0.3 \xt{122} k$, dwarfing the entropy of its interior, $S_{\mathrm{CEH\ int}} = 1.2^{+1.1}_{-0.7}\xt{103} k$.
	We make the first tentative estimate of the entropy of weakly interacting massive particle dark 
	matter within the observable universe, $S_{\mathrm{dm}} = 10^{88\pm1} k$. 
	We highlight several caveats pertaining to these estimates and make recommendations for future work.
\end{abstract}


\maketitle

\section{Introduction} \label{introduction}

The entropy budget of the universe is important because its increase is associated
with all irreversible processes, on all scales, across all 
facets of nature: gravitational clustering, accretion disks, supernovae, stellar fusion, 
terrestrial weather, and chemical, geological and biological processes 
\citep{Frautschi1982,LineweaverEgan2008}.

Recently, \citet{Frampton2008} and \citet{Frampton2008b} 
reported the entropy budget of the observable universe. Their budgets (listed aside 
others in Table \ref{tab:currententropy}) estimate the total entropy of the observable 
universe to be $S_{\mathrm{obs}} \sim 10^{102} k - 10^{103} k$, dominated by the entropy of 
supermassive black holes (SMBHs) at the centers of galaxies.  
That the increase of entropy has not yet been capped by some limiting value, 
such as the holographic bound \citep{tHooft1993,Susskind1995} at 
$S_{\mathrm{max}} \sim 10^{123} k$ \citep{Frampton2008}, is the reason dissipative 
processes are ongoing and that life can exist. 

In this paper, we improve the entropy budget by using recent observational data
and quantifying uncertainties.
The paper is organized as follows. In what remains of the Introduction, we describe 
two different schemes for quantifying the increasing entropy of the universe, and we 
comment on caveats involving the identification of gravitational entropy.
Our main work is presented in Sections \ref{matter} and  \ref{scheme2}, where we 
calculate new entropy budgets within each of the two accounting schemes. 
We finish in Section \ref{discussion} with a discussion touching on the time evolution 
of the budgets we have calculated, and ideas for future work.

Throughout this paper we assume flatness ($\Omega_k = 0$) as predicted by 
inflation \citep{Guth1981,Linde1982} and supported by observations 
\citep{Spergel2007}. Adopted values for other cosmological parameters are 
$h = 0.705 \pm 0.013$, 
$\omega_b = \Omega_b h^2 = 0.0224 \pm 0.0007$,
$\omega_m = \Omega_m h^2 = 0.136 \pm 0.003$ \citep{Seljak2006}, 
and
$T_{\mathrm{CMB}} = 2.725 \pm 0.002\ K$ (\citealt{Mather1999}; quoted uncertainties are $1\sigma$).


\subsection{Two Schemes for Quantifying the Increasing Entropy of the Universe} \label{frwgsl}

Modulo statistical fluctuations, the generalized second law of thermodynamics 
holds that the entropy of the universe (including Bekenstein-Hawking entropy in 
the case of any region hidden behind an event horizon), must not decrease 
with time \citep{Bekenstein1974,Gibbons1977}. Within the FRW framework, the 
generalized second law can be applied in at least two obvious ways: 
\begin{enumerate}
	\item The total entropy in a sufficiently large comoving volume of the universe 
		does not decrease with cosmic time,
		\begin{equation}
			d S_{\mathrm{comoving\ volume}} \ge 0. \label{eq:one}
		\end{equation}
	\item The total entropy of matter contained within the cosmic event horizon 
		(CEH) plus the entropy of the CEH itself, does not decrease with cosmic 
		time,
		\begin{equation}
			d S_{\mathrm{CEH\ interior}} + d S_{\mathrm{CEH}} \ge 0. \label{eq:two}
		\end{equation}
\end{enumerate}

In the first of these schemes, the system is bounded by a closed comoving surface.
The system is effectively isolated because large-scale homogeneity and 
isotropy imply no net flows of entropy into or out of the comoving volume. The 
time-slicing in this scheme is along surfaces of constant cosmic time. 
Event horizons of black holes are used to quantify the entropy of black holes, 
however the CEH is neglected since the assumption of 
large-scale homogeneity makes it possible for us to keep track of the entropy of matter 
beyond it. A reasonable choice for the comoving volume in this scheme is the comoving 
sphere that presently corresponds to the observable universe, i.e., the 
\ifthenelse{\boolean{colver}} { 
gray area in Figure \ref{fig:FRW}. }{
dark gray area in Figure \ref{fig:FRW}. }
Correspondingly, in Section \ref{matter} we calculate the present entropy budget of 
the observable universe and we do not include the CEH.

The second scheme is similar to the first in that we time-slice along surfaces of 
constant cosmic time. However, here the system 
\ifthenelse{\boolean{colver}} { 
(yellow shade in Figure \ref{fig:FRW}) }{
(light gray shade in Figure \ref{fig:FRW}) }
is bounded by the time-dependent CEH instead of a comoving 
boundary. Migration of matter across the CEH is not negligible, and the CEH 
entropy \citep{Gibbons1977} must be included in the budget to account for 
this (e.g.\ \citealt{Davis2003}). The present entropy of the CEH 
and its interior is calculated in Section \ref{scheme2}.

%
%
\ifthenelse{\boolean{colver}} {
	\begin{figure}[!hbtp]
       		\begin{center}
               		\includegraphics[width=\linewidth]{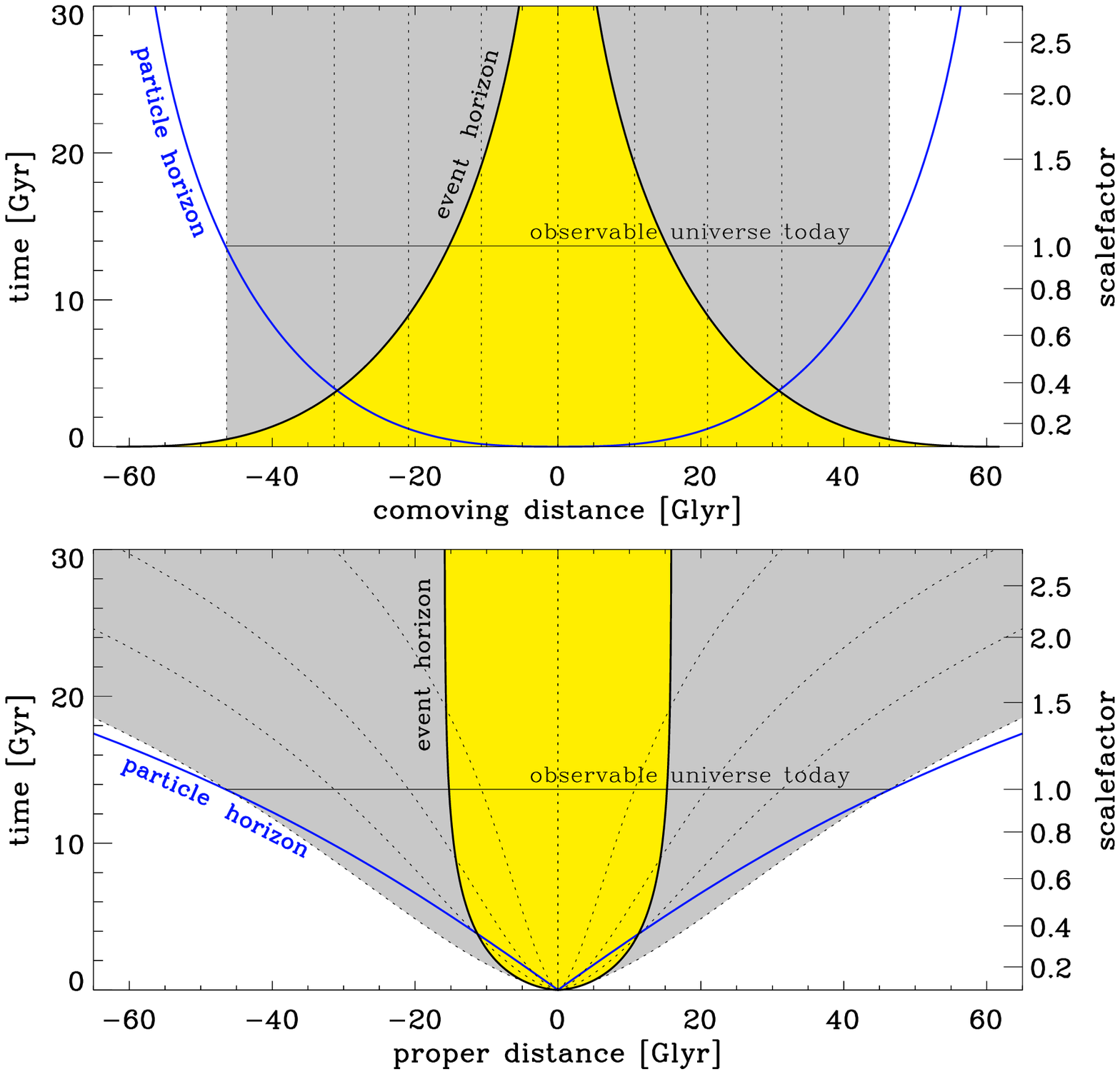}
               		\caption{These two panels show the particle horizon (see Equation \ref{eq:robs} 
               		and Figure \ref{fig:dehh0}) and the cosmic event horizon (see Equation \ref{eq:deh}) 
               		as a function of time. The difference between the two panels is the
               		spatial coordinate system used: the $x$-axis in the bottom panel is proper
               		distance $D$ and in the top panel it is comoving distance $\chi \equiv \frac{D}{a}$,
               		where $a$ is the cosmic scalefactor.
               		The origin is chosen so that our galaxy is the central vertical dotted line.
               		The other dotted lines represent distant galaxies, which are approximately 
               		comoving and recede as the universe expands. The region inside the 
			particle horizon is the observable universe. The comoving volume that 
			corresponds to the observable universe today, about $13.7\ Gyr$ 
               		after the big bang, is filled gray. In scheme 1, the entropy within this 
               		comoving volume increases (or remains constant) with time. 
               		Alternatively, in scheme 2 the entropy within the event horizon (the region 
               		filled yellow), plus the entropy of the horizon itself, increases (or remains 
               		constant) with time.}
               		\label{fig:FRW}
       		\end{center}
	\end{figure}
}{
	\begin{figure}[!hbtp]
       		\begin{center}
               		\includegraphics[width=\linewidth]{frw_top_two_panels_bw.ps}
               		\caption{These two panels show the particle horizon (see Equation \ref{eq:robs} 
               		and Figure \ref{fig:dehh0}) and the cosmic event horizon (see Equation \ref{eq:deh}) 
               		as a function of time. The difference between the two panels is the
               		spatial coordinate system used: the $x$-axis in the bottom panel is proper
               		distance $D$ and in the top panel it is comoving distance $\chi \equiv \frac{D}{a}$,
               		where $a$ is the cosmic scalefactor.
               		The origin is chosen so that our galaxy is the central vertical dotted line.
               		The other dotted lines represent distant galaxies, which are approximately 
               		comoving and recede as the universe expands. The region inside the 
			particle horizon is the observable universe. The comoving volume that 
			corresponds to the observable universe today, about $13.7\ Gyr$ 
               		after the big bang, is filled dark gray. In scheme 1, the entropy within this 
			comoving volume increases (or remains constant) with time. 
               		Alternatively, in scheme 2 the entropy within the event horizon (the light gray 
			region), plus the entropy of the horizon itself, increases (or remains 
               		constant) with time. \textit{[See the electronic edition of the journal for a 
			 version of this figure.]}}
               		\label{fig:FRW}
       		\end{center}
	\end{figure}
}

\subsection{Entropy and Gravity}

It is widely appreciated that non-gravitating systems of particles evolve 
toward homogenous temperature and density distributions. The corresponding 
increase in the volume of momentum-space and position-space occupied by the
constituent particles represents an increase in entropy.
On the other hand, strongly 
gravitating systems become increasingly lumpy. With ``lumpyness'' naively akin to 
``orderliness'', it is not as easy to see that the total entropy increases. In these systems
the entropy is shared among numerous components, all of which must be considered.



For example, approximately collisionless long-range gravitational interactions between
stars result in dynamical relaxation of galaxies (whereby bulk motions are dissipated and 
entropy is transferred to stars in the outer regions of the galaxy; \citealt{LyndenBell1967}) 
and stellar evaporation from galaxies (whereby stars are ejected altogether, carrying with 
them energy, angular momentum and entropy, and allowing what remains behind to 
contract; e.g.\ \citealt{Binney2008}). 
In more highly dissipative systems, i.e., accretion disks, non-gravitational 
interactions (viscosity and/or magnetorotational instability; \citealt{Balbus2002}) transfer 
angular momentum and dissipate energy and entropy.

In addition to these considerations, entropy also increases when gravitons are produced. 
A good example is the in-spiral of close binaries, such as the Hulse-Taylor 
binary pulsar system \citep{HulseTaylor1975,Weisberg2005}. Gravitational waves
emitted from the system extract orbital energy (and therefore entropy) allowing the system 
to contract. 

The entropy of a general gravitational field is still not known. 
\citet{Penrose1987,Penrose1979,Penrose2004}
has proposed that it is related to the Weyl curvature tensor $W_{\mu \nu \kappa \lambda}$. 
In conformally flat spacetimes (such as an ideal FRW universe), the Weyl curvature vanishes 
and gravitational entropy is postulated to vanish (to limits imposed by quantum uncertainty). In 
clumpy spacetimes the Weyl curvature takes large values and the gravitational entropy is high. 
While Ricci curvature $R_{\mu \nu}$ vanishes in the absence of matter, Weyl curvature may 
still be non-zero (e.g.\ gravitational waves traveling though empty space) and the corresponding 
gravitational entropy may be non-zero.

If these ideas are correct then the low gravitational entropy of the early universe comes from 
small primordial gravitational perturbations. Gravitational entropy then increases with the growing 
amplitude of linear density fluctuations parameterized through the matter power spectrum $P(k)$. 
The present gravitational entropy, however, is expected to be dominated by the nonlinear 
overdensities (with large Weyl tensors) which have formed since matter-radiation equality. 

In extreme cases, gravitational clumping leads to the formation of black holes. The
entropy of black holes is well known \citep{Bekenstein1973,Hawking1976,Strominger1996}. 
The entropy of a Schwarzschild black hole is given by 
\begin{eqnarray}
	S_ {\mathrm{BH}} = \frac{k c^3}{G \hbar} \frac{A}{4}  = \frac{4 \pi k G}{c \hbar} M^2 \label{eq:areaent}
\end{eqnarray}
where $A = \frac{16 \pi G^2 M^2}{c^4}$ is the event-horizon area and $M$ is the black hole
mass. 



Because gravitational entropy is difficult to quantify, we only include it in the two extremes:
the thermal distribution of gravitons and black holes.



\section{The Present Entropy of the Observable Universe} \label{matter} \label{scheme1}

The present entropy budget of the observable universe was estimated most recently by 
\citet{Frampton2008} and \citet{Frampton2008b}. Those papers and earlier work 
\citep{Kolb1981,Frautschi1982,Penrose2004,Bousso2007} identified the largest contributors 
to the entropy of the observable universe as black holes, followed distantly by the cosmic 
microwave background (CMB) and the neutrino background. The last column of Table 
\ref{tab:currententropy} contains previous estimates of the entropy in black holes, the CMB and 
neutrinos, as well as several less significant components.

Sections \ref{bar} -- \ref{smbh} below describe the data and assumptions 
used to calculate our entropy densities (given in Column 2 of Table \ref{tab:currententropy}). 
Our entropy budget for the observable universe (Column 3 of Table \ref{tab:currententropy}) 
is then found by multiplying the entropy density by the volume of the observable 
universe $V_{\mathrm{obs}}$,
\begin{eqnarray}
	S_{i} & = & s_{i} V_{\mathrm{obs}}
\end{eqnarray}
where $s_i$ is the entropy density of component $i$. The volume of the observable 
universe is (see Appendix)
\begin{eqnarray}
	V_{\mathrm{obs}} & = & 43.2 \pm 1.2 \xt{4}\ {Glyr}^3 \nonumber \\
		& = & 3.65 \pm 0.10 \xt{80}\ m^3.
\end{eqnarray}




\begin{deluxetable*}{l l l r}
\tablecaption{Current Entropy of the Observable Universe (Scheme 1 Entropy Budget) \label{tab:currententropy}}
\tablehead{
	\colhead{Component} &
	\colhead{Entropy Density $s$ $[k\ m^{-3}]$} &
	\colhead{Entropy $S$ $[k]$} &
	\colhead{Entropy $S$ $[k]$ (Previous Work)}
}
\startdata
	SMBHs
		& $8.4^{+8.2}_{-4.7} \xt{23}$
		& $3.1^{+3.0}_{-1.7} \xt{104}$
		& $10^{101}[1]$, $10^{102}[2]$, $10^{103}[3]$
		\\
	Stellar BHs $(2.5 - 15\ M_{\odot})$
		& $1.6\xt{17^{+0.6}_{-1.2}}$
		& $5.9\xt{97^{+0.6}_{-1.2}}$		
		& $10^{97}[2]$, $10^{98}[4]$
		\\
	Photons
		& $1.478\pm0.003\xt{9}$
		& $5.40\pm0.15\xt{89}$				
		& $10^{88}[1,2,4]$, $10^{89}[5]$	
		\\
	Relic Neutrinos
		& $1.411\pm0.014\xt{9}$
		& $5.16\pm0.15\xt{89}$
		& $10^{88}[2]$,$10^{89}[5]$		
		\\	
	\addthis{WIMP} Dark Matter
		& $5\xt{7 \pm 1}$
		& $2\xt{88 \pm 1}$	
		& $-$												
		\\
	Relic Gravitons
		& $1.7\xt{7^{+0.2}_{-2.5}}$
		& $6.2\xt{87^{+0.2}_{-2.5}}$				
		& $10^{86}[2,3]$												
		\\
	ISM and IGM
		& $20 \pm 15$ 
		& $7.1 \pm 5.6 \xt{81}$	
		& $-$												
		\\
	Stars
		& $0.26 \pm 0.12$  
		& $9.5 \pm 4.5 \xt{80}$	
		& $10^{79}[2]$
		\\
	\hline
	{\bf Total}
		& \boldmath$8.4^{+8.2}_{-4.7}\xt{23}$
		& \boldmath$3.1^{+3.0}_{-1.7}\xt{104}$ 	
		& \boldmath$10^{101}[1]$, $10^{102}[2]$, $10^{103}[3]$
		\\
	\hline
	\addthis{Tentative Components:}
		&
		&
		&
		\\
	\addthis{Massive Halo BHs $(10^5\ M_{\odot})$}
		& \addthis{$10^{25}$}
		& \addthis{$10^{106}$}	
		& \addthis{$10^{106}[6]$}
		\\
	\addthis{Stellar BHs $(42 - 140\ M_{\odot})$}
		& \addthis{$8.5\xt{18^{+0.8}_{-1.6}}$}
		& \addthis{$3.1\xt{99^{+0.8}_{-1.6}}$}
		& \addthis{$-$}
\enddata
\tablecomments{
	Our budget is consistent with previous estimates from the literature with the exception that
	SMBHs, which dominate the budget, contain at least an order of magnitude more 
	entropy as previously estimated, due to the contributions of black holes $100$ times larger
	than those considered in previous budgets.
	Uncertainty in the volume of the observable universe (see Appendix) has been included in the quoted 
	uncertainties. \removethis{Stellar black holes in the mass range $42-140\ M_{\odot}$ (marked with an 
	$*$) are included tentatively since their existence is speculative.} \addthis{Massive halo black holes at 
	$10^5\ M_{\odot}$ and stellar black holes in the range $42-140\ M_{\odot}$ are included tentatively 
	since their existence is speculative. They are not counted in the budget totals.}
	Previous work: 
	[1] \citet{Penrose2004},
	[2] \citet{Frampton2008},
	[3] \citet{Frampton2008b},
	[4] \citet{Frautschi1982},
	[5] \citet{Kolb1981},
	\addthis{[6] \citet{Frampton2009b}.}
	}
\end{deluxetable*}

\subsection{Baryons} \label{bar}

%

For a non-relativistic, non-degenerate gas the specific entropy (entropy per baryon) is given 
by the Sakur-Tetrode equation (e.g.\ \citealt{Basu1990})
\begin{eqnarray}
	(s/n_b) = \frac{k}{n_b} \sum_i n_i \ln \left[ Z_i(T) (2 \pi m_i k T)^{\frac{3}{2}} e^{\frac{5}{2}} n_i^{-1} h^{-3} \right], \label{eq:sakurtetrode}
\end{eqnarray}
where $i$ indexes particle types in the gas, $n_i$ is the $i$th particle type's number density, 
and $Z_i(T)$ is its internal partition function. \citet{Basu1990} found specific entropies between $11\ k$ 
and $21\ k$ per baryon for main-sequence stars of approximately solar mass. For components of the 
interstellar medium (ISM) and intergalactic medium (IGM) they found specific entropies between 
$20\ k$ ($H_{2}$ in the ISM) and $143\ k$ (ionized hydrogen in the IGM) per baryon. 

The cosmic entropy density in stars $s_{*}$ can be estimated by multiplying the specific entropy 
of stellar material by the cosmic number density of baryons in stars $n_{b*}$:
\begin{eqnarray}
	s_{*} = (s/n_b)_{*} n_{b*} = (s/n_b)_{*} \frac{\rho_{*}}{m_{p}} = (s/n_b)_{*} \left[ \frac{3 H^2}{8 \pi G} \frac{\Omega_{*}}{m_p} \right]. 
\end{eqnarray}
Using the stellar cosmic density parameter $\Omega_{*} = 0.0027 \pm 0.0005$ \citep{Fukugita2004}, 
and the range of specific entropies for main-sequence stars around the solar mass 
(which dominate stellar mass), we find 
\begin{eqnarray}
	%
	%
	s_{*} & = & 0.26\pm0.12\ k\ m^{-3}, \\
	S_{*} & = & 9.5 \pm 4.5 \xt{80}\ k. \label{eq:sstars}
\end{eqnarray}
Similarly, the combined energy density for the ISM and IGM is $\Omega_{\mathrm{gas}} = 0.040 \pm 0.003$ 
\citep{Fukugita2004}, and by using the range of specific entropies for ISM and IGM components, we 
find
\begin{eqnarray}
	%
	%
	s_{\mathrm{gas}} & = & 20\pm{15}\ k\ m^{-3}, \\
	S_{\mathrm{gas}} & = & 7.1\pm 5.6 \xt{81}\ k. \label{eq:sismigm}
\end{eqnarray}
The uncertainties in Equations (\ref{eq:sstars}) and (\ref{eq:sismigm}) are dominated by uncertainties in the
mass weighting of the specific entropies, but also include uncertainties in 
$\Omega_{*}$, $\Omega_{\mathrm{gas}}$ and the volume of the observable universe. 

\subsection{Photons} \label{bpg}

The CMB photons are the most significant non-black hole 
contributors to the entropy of the observable universe. The distribution of CMB photons is 
thermal \citep{Mather1994} with a present temperature of 
$T_{\gamma} = 2.725 \pm 0.002\ K$ \citep{Mather1999}.

The entropy of the CMB is calculated using the equation for a black body (e.g.\ \citet{Kolb1990}),
\begin{eqnarray}
	s_{\gamma} 	& = & \frac{2 \pi^2}{45} \frac{k^4}{c^3 \hbar^3} g_{\gamma} T_{\gamma}^{3} \label{eq:radent} \\
				& = & 1.478 \pm 0.003 \xt{9}\ k\ m^{-3}, \nonumber \\
	S_{\gamma}	& = & 2.03 \pm 0.15 \xt{89}\ k, \label{eq:sgamma}
\end{eqnarray}
where $g_{\gamma} = 2$ is the number of photon spin states. The uncertainty in Equation (\ref{eq:sgamma}) 
is dominated by uncertainty in the size of the observable universe.

The non-CMB photon contribution to the entropy budget (including starlight and heat emitted
by the ISM) is somewhat less, at around $10^{86} k$ \citep{Frautschi1982,Bousso2007,Frampton2008}. 

\subsection{Relic Neutrinos} \label{nusec}


The neutrino entropy cannot be calculated directly since the temperature of cosmic neutrinos has
not been measured. Standard treaties of the radiation era (e.g.\ \citealt{Kolb1990,Peacock1999}) 
describe how the present temperature (and entropy) of massless relic neutrinos can be calculated 
from the well known CMB photon temperature. Since this background physics is required 
for Sections \ref{relicgrav} and \ref{darkmatter}, we summarize it briefly here.

A simplifying feature of the radiation era (at least at known energies $\lsim 10^{12} eV$) is that the 
radiation fluid evolves adiabatically: the entropy density decreases as the cube of the increasing 
scalefactor $s_{\mathrm{rad}} \propto a^{-3}$. The evolution is adiabatic because reaction rates in the fluid 
are faster than the expansion rate $H$ of the universe. It is convenient to write the entropy density as
\begin{eqnarray}
	s_{\mathrm{rad}} & = & \frac{2 \pi^2}{45} \frac{k^4}{c^3 \hbar^3} g_{*S} T_{\gamma}^{3} \propto a^{-3} \label{eq:srad}
\end{eqnarray}
where $g_{*S}$ is the number of relativistic degrees of freedom in the fluid (with $m < kT/c^2$) 
given approximately by
\begin{eqnarray}
	g_{*S}(T) & \approx & \sum_{
	{\scriptsize \begin{array}{c}
	bosons,\ i
	\end{array}}
	} g_{i} \left( \frac{T_{i}}{T_{\gamma}} \right)^{3} + \sum_{
	{\scriptsize \begin{array}{c}
	fermions,\ j
	\end{array}}
	} \frac{7}{8} g_{j} \left( \frac{T_{j}}{T_{\gamma}} \right)^{3}.
\end{eqnarray}
For photons alone, $g_{*S} = g_{\gamma} = 2$, and thus Equation (\ref{eq:srad}) becomes Equation (\ref{eq:radent}). 
For photons coupled to an electron-positron component, such as existed before electron-positron 
annihilation,
$g_{*S} = g_{\gamma} + \frac{7}{8} g_{e\pm} = 2 + \frac{7}{8} 4 = \frac{11}{2}$. 

As the universe expands, massive particles annihilate, heating the
remaining fluid. The effect on the photon temperature is quantified by inverting Equation (\ref{eq:srad}), 
\begin{eqnarray}
	T_{\gamma} 
				& \propto & a^{-1} g_{*S}^{-1/3}.  \label{eq:gheat}
\end{eqnarray}
The photon temperature decreases less quickly than $a^{-1}$ because $g_{*S}$ decreases 
with time. 
Before electron-positron $e^{\pm}$ annihilation the temperature of the photons was the same 
as that of the almost completely decoupled neutrinos. After $e^{\pm}$ annihilation, heats only 
the photons, the two temperatures differ by a factor $C$,
\begin{eqnarray}
	T_{\nu} = C\ T_{\gamma}. \label{eq:nucool}
\end{eqnarray}
A reasonable approximation $C \approx (4/11)^{1/3}$ is derived by assuming that only
photons were heated during $e^{\pm}$ annihilation, where $4/11$ is the ratio of 
$g_{*S}$ for photons to $g_{*S}$ for photons, electrons, and positrons. 

Corrections are necessary at the $10^{-3}$ level because neutrinos had not completely
decoupled at $e^{\pm}$ annihilation \citep{Gnedin1998}. 
The neutrino entropy density is computed assuming a thermal distribution with 
$T_{\nu} = (4/11)^{1/3} T_{\gamma}$, and we assign a $1\%$ uncertainty. 


\begin{eqnarray}
	s_{\nu} 	& = & \frac{2 \pi^2}{45} \frac{k^4}{c^3 \hbar^3} g_{\nu} \left( \frac{7}{8} \right) T_{\nu}^{3} \nonumber \\
			& = & 1.411 \pm 0.014 \xt{9}\ k\ m^{-3}, \label{eq:nu0ent}
\end{eqnarray}	
where $g_{\nu} = 6$ ($3$ flavors, $2$ spin states each).  The total neutrino entropy in the 
observable universe is then
\begin{eqnarray}
	S_{\nu}	& = & 5.16 \pm 0.14 \xt{89}\ k \label{eq:nuentobs}
\end{eqnarray}
with an uncertainty dominated by uncertainty in the volume of the observable universe.

Neutrino oscillation experiments have demonstrated that neutrinos are massive by measuring 
differences between the three neutrino mass eigenstates \citep{Cleveland1998,Adamson2008,Abe2008}. 
At least two of the mass eigenstates are heavier than $\sim 0.009\ eV$. 
Since this is heavier than their current relativistic energy ($\frac{k}{2}\ C\ T_{\gamma} = 0.0001\ eV$; 
computed under the assumption that they are massless) at least two of the three masses are presently 
non-relativistic. 

Expansion causes non-relativistic species to cool as $a^{-2}$ instead of $a^{-1}$, which would
result in a lower temperature for the neutrino background than suggested by Equation (\ref{eq:nucool}). 
The entropy density (calculated in Equation \ref{eq:nu0ent}) and entropy (calculated in Equation \ref{eq:nuentobs}) are 
unaffected by the transition to non-relativistic cooling since the cosmic expansion of relativistic and 
non-relativistic gases are both adiabatic processes (the comoving entropy is conserved, so in either 
case $s \propto a^{-3}$).


We neglect a possible increase in neutrino entropy due to their infall into gravitational potentials 
during structure formation. If large, this will need to be considered in future work.

\subsection{Relic Gravitons} \label{relicgrav}


A thermal background of gravitons is expected to exist, which decoupled from 
the photon bath around the Planck time, and has been cooling as $T_{\mathrm{grav}} \propto a^{-1}$ 
since then. The photons cooled less quickly because they were heated by the annihilation 
of heavy particle species (Equation \ref{eq:gheat}). Thus we can relate the current graviton 
temperature to the current photon temperature
\begin{eqnarray}
	T_{\mathrm{grav}} & = & \left(\frac{g_{*S}(t_0)}{g_{*S}(t_{\mathrm{planck}})}\right)^{1/3} T_{\gamma}, \label{eq:gravtemp1} 
\end{eqnarray}
where $g_{*S}(t_{\mathrm{planck}})$ is the number of relativistic degrees of freedom at the Planck
time and $g_{*S}(t_0) = 3.91$ today (this is appropriate even in the case of massive neutrinos
because they decoupled from the photon bath while they were still relativistic). Given the 
temperature of background gravitons, their entropy can be calculated as 
\begin{eqnarray}
	s_{\mathrm{grav}} 	& = & \frac{2 \pi^2}{45} \frac{k^4}{c^3 \hbar^3} g_{\mathrm{grav}} T_{\mathrm{grav}}^{3} \label{eq:gravent}
\end{eqnarray}
where $g_{\mathrm{grav}} = 2$. 

Figure \ref{fig:gstar} shows $g_{*S}$ as a function of temperature. The function is well 
known for temperatures below about $10^{12} eV$, but is not known at higher 
temperatures. Previous estimates of the background graviton entropy have assumed 
$g_{*S}(t_{\mathrm{planck}}) \sim g_{*S}(10^{12} eV) = 106.75$ \citep{Frampton2008,Frampton2008b}, 
but this should be taken as a lower bound on $g_{*S}(t_{\mathrm{planck}})$ yielding an upper 
bound on $T_{\mathrm{grav}}$ and $s_{\mathrm{grav}}$.

To get a better idea of the range of possible graviton temperatures and entropies, we 
have adopted three values for $g_{*S}(t_{\mathrm{planck}})$. As a minimum likely value we use
$g_{*S}=200$ 
\ifthenelse{\boolean{colver}} { 
(Figure \ref{fig:gstar}, thick blue line), }{
(Figure \ref{fig:gstar}, solid gray line), }
which includes the minimal set of additional particles suggested by 
supersymmetry. As our middle value we use $g_{*S}=350$, corresponding to the 
linear extrapolation of $g_{*S}$ in $\log(T)$ to the Planck scale 
\ifthenelse{\boolean{colver}} { 
(Figure \ref{fig:gstar}, gray line). }{
(Figure \ref{fig:gstar}, dotted gray line). }
And as a maximum likely value we use $g_{*S}=10^{5}$, corresponding to an exponential 
extrapolation 
\ifthenelse{\boolean{colver}} { 
(Figure \ref{fig:gstar}, thin blue line). }{
(Figure \ref{fig:gstar}, dashed gray line). }
\ifthenelse{\boolean{colver}} {
	\begin{figure}[!hbtp]
       		\begin{center}
               		\includegraphics[width=\linewidth]{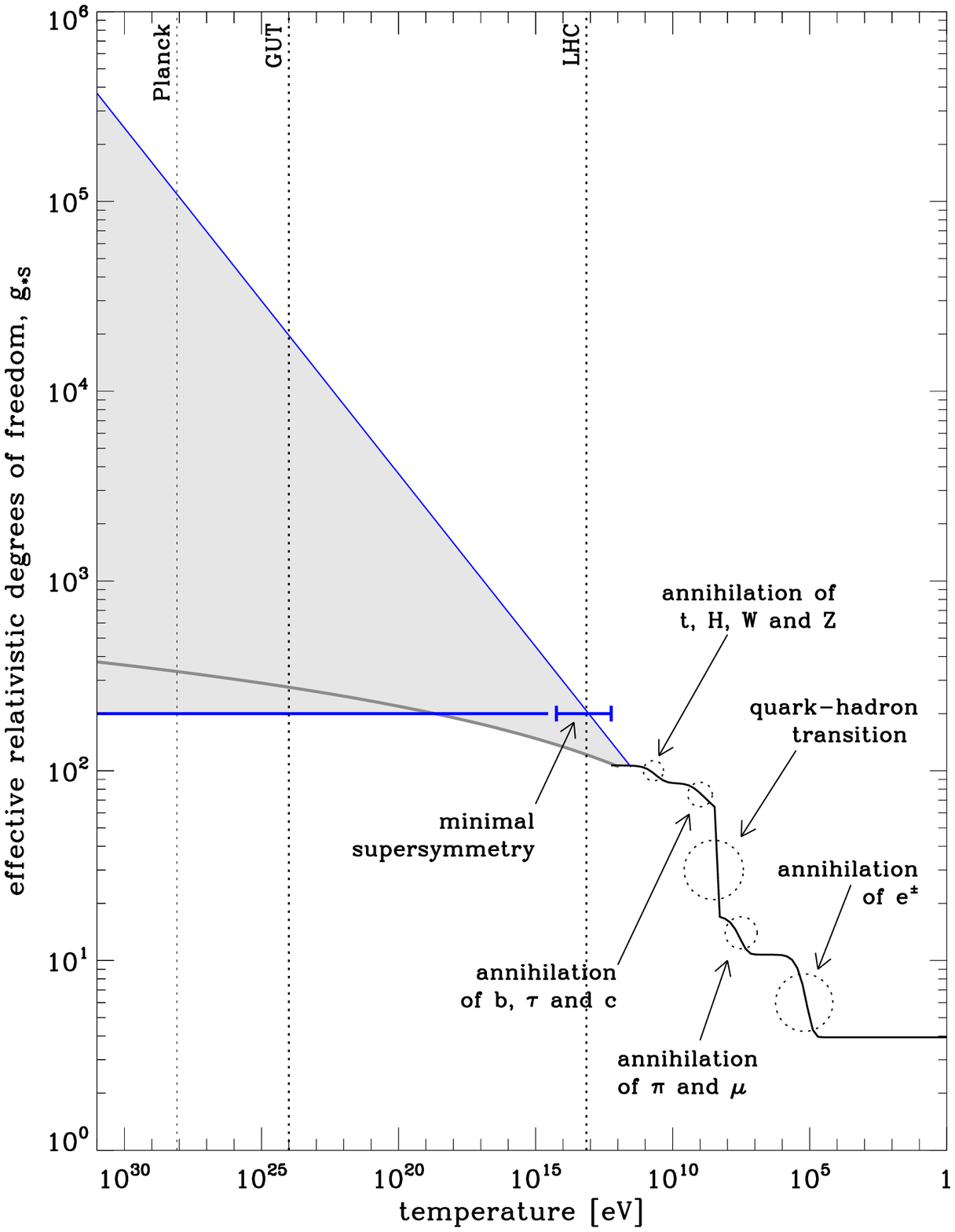}
               		\caption{Number of relativistic degrees of freedom $g_{*S}$ as a function
               		of temperature, computed using the prescription given by \citet{Coleman2003}. 
               		All the particles of the standard model are relativistic at $T \gsim 10^{12}\ eV$ 
               		and $g_{*S}(10^{12}\ eV)=106.75$. The value of $g_{*S}$ is not known above $T \sim 10^{12}$. 
               		To estimate plausible ranges of values, we extrapolate $g_{*S}$ linearly (gray
			line) and exponentially (thin blue line) in $\log(T)$. The minimum contribution 
               		to $g_{*S}$ from supersymmetric partners is shown (blue bar) and taken to 
			indicate a minimum likely value of $g_{*S}$ at higher temperatures (thick blue line).
              		}
               		\label{fig:gstar}
       		\end{center}
	\end{figure}
}{
	\begin{figure}[!hbtp]
       		\begin{center}
               		\includegraphics[width=\linewidth]{g_star_predictor_pretty_log_bw.ps}
               		\caption{Number of relativistic degrees of freedom $g_{*S}$ as a function
               		of temperature, computed using the prescription given by \citet{Coleman2003}. 
               		All the particles of the standard model are relativistic at $T \gsim 10^{12}\ eV$ 
               		and $g_{*S}(10^{12}\ eV)=106.75$. The value of $g_{*S}$ is not known above $T \sim 10^{12}$. 
               		To estimate plausible ranges of values, we extrapolate $g_{*S}$ linearly (dotted 
               		line) and exponentially (dashed line) in $\log(T)$. The minimum contribution 
               		to $g_{*S}$ from supersymmetric partners is shown and taken to indicate a 
               		minimum likely value of $g_{*S}$ at higher temperatures. \textit{[See the electronic 
			edition of the journal for a color version of this figure.]}}
               		\label{fig:gstar}
       		\end{center}
	\end{figure}
}

%
%
%
The corresponding graviton temperatures today are (Equation \ref{eq:gravtemp1})
\begin{eqnarray}
		T_{\mathrm{grav}} & = & 0.61^{+0.12}_{-0.52} \textrm{ K}.
\end{eqnarray}
Inserting this into Equation (\ref{eq:gravent}) we find the entropy in the relic graviton 
background to be 
\begin{eqnarray}
	s_{\mathrm{grav}}	& = & 1.7 \xt{7^{+0.2}_{-2.5}}\ k\ m^{-3},  \\
	S_{\mathrm{grav}}	& = & 6.2 \xt{87^{+0.2}_{-2.5}}\ k.
\end{eqnarray}


It is interesting to note the possibility of applying Equation (\ref{eq:gravtemp1}) in 
reverse, i.e., calculating the number of relativistic degrees of freedom at the Planck time 
using future measurements of the graviton background temperature.

\subsection{Dark Matter} \label{darkmatter}


The most compelling interpretation of dark matter is as a weakly interacting superpartner
(or weakly interacting massive particle, WIMP). 
According to this idea, dark matter particles decoupled from the radiation background at some 
energy above the particle mass. 

If this interpretation is correct, the fraction of relativistic background entropy in dark matter 
at the time dark matter decoupled $t_{\mathrm{dm\ dec}}$ is determined by the fraction of relativistic 
degrees of freedom that were associated with dark matter at that time (see Equation \ref{eq:srad}). 
\begin{eqnarray}
	s_{\mathrm{dm}} = \frac{g_{*S\ \mathrm{dm}}(t_{\mathrm{dm\ dec}})}{g_{*S\ \mathrm{non-dm}}(t_{\mathrm{dm\ dec}})} s_{\mathrm{non-dm\ rad}} \label{eq:sdm}
\end{eqnarray}
This can be evaluated at dark matter decoupling, or any time thereafter, since both $s_{\mathrm{dm}}$
and $s_{\mathrm{non-dm\ rad}}$ are adiabatic ($\propto a^{-3}$).



We are unaware of any constraint on the number of superpartners that may collectively 
constitute dark matter. The requirements that they are only weakly interacting, and that they 
decouple at a temperature above their mass, are probably only satisfied by a few (even one) 
species. Based on these arguments, we assume $g_{*S\ dm}(t_{\mathrm{dm\ dec}}) \lsim 20$ and 
$g_{*S}(t_{\mathrm{dm\ dec}}) \gsim 106.75$ which yields the upper limit 
\begin{eqnarray}
	\frac{g_{*S\ \mathrm{dm}}(t_{\mathrm{dm\ dec}})}{g_{*S}(t_{\mathrm{dm\ dec}})} \lsim \frac{1}{5}. \label{eq:sdmhi}
\end{eqnarray}

On the other hand there may be many more degrees of freedom than suggested by minimal 
supersymmetry. By extrapolating $g_{*S}$ exponentially beyond supersymmetric scales 
(to $10^{15}\ eV$), we find $g_{*S}(t_{\mathrm{dm\ dec}}) \lsim 800$. In the simplest case, dark matter 
is a single scalar particle so $g_{*S\ \mathrm{dm}}(t_{\mathrm{dm\ dec}}) \gsim 1$ and we take as 
a lower limit
\begin{eqnarray}
	\frac{g_{*S\ \mathrm{dm}}(t_{\mathrm{dm\ dec}})}{g_{*S\ \mathrm{non-dm}}(t_{\mathrm{dm\ dec}})} \gsim \frac{1}{800}. \label{eq:sdmlo}
\end{eqnarray}



Inserting this into Equation (\ref{eq:sdm}) at the present day gives 
\begin{eqnarray}
	s_{\mathrm{dm}} & = & 5\xt{7\pm1}\ k\ m^{-3},
\end{eqnarray}
where we have used the estimated limits given in Equations (\ref{eq:sdmhi}) and (\ref{eq:sdmlo}) 
and taken $s_{\mathrm{non-dm\ rad}}$ to be the combined entropy of neutrinos and radiation today
(Equations \ref{eq:radent} and \ref{eq:nu0ent}). The corresponding estimate for the total dark matter 
entropy in the observable universe is
\begin{eqnarray}
	S_{\mathrm{dm}} & = & 2\xt{88\pm1}\ k.
\end{eqnarray}

As with our calculated neutrino entropy, our estimates here carry the caveat that we have not 
considered changes in the dark matter entropy associated with gravitational structure formation.

\subsection{Stellar Black Holes}

In the top panel of Figure \ref{fig:sbhentropy} we show the stellar initial mass function (IMF) 
parameterized by
\begin{eqnarray}
	\frac{d n_{\mathrm{initial}}}{d \log(M)} \propto \left(\frac{M}{M_{\odot}}\right)^{\alpha + 1},
\end{eqnarray}
with $\alpha = -1.35$ at $M < 0.5 M_{\odot}$ and $\alpha = -2.35^{+0.65}_{-0.35}$ at $M \ge 0.5 M_{\odot}$
\citep{Elmegreen2007}. We also show the present distribution of main-sequence stars, 
which is proportional to the initial distribution for $M \lsim 1 M_{\odot}$,
but which is reduced by a factor of $(M/M_{\odot})^{-2.5}$ for heavier stars \citep{Fukugita2004}. 
\begin{equation}
	\frac{d n_{\mathrm{present}}}{d \log(M)} = \left\{ 
	\begin{array}{ll} 
		\frac{d n_{\mathrm{initial}}}{d \log(M)}, 								& \mbox{for $M < 1 M_{\odot}$} \\
		\frac{d n_{\mathrm{initial}}}{d \log(M)} \left( \frac{M}{M_{\odot}} \right)^{-2.5}, 	& \mbox{for $M \ge 1 M_{\odot}$}
	\end{array}
	\right. .
\end{equation}
The initial and present distributions are normalized using the present cosmic density of 
stars, $\Omega_* = 0.0027 \pm 0.0005$ \citep{Fukugita2004}.
\ifthenelse{\boolean{colver}} {
	\begin{figure}[!hbtp]
       		\begin{center}
               		\includegraphics[width=\linewidth]{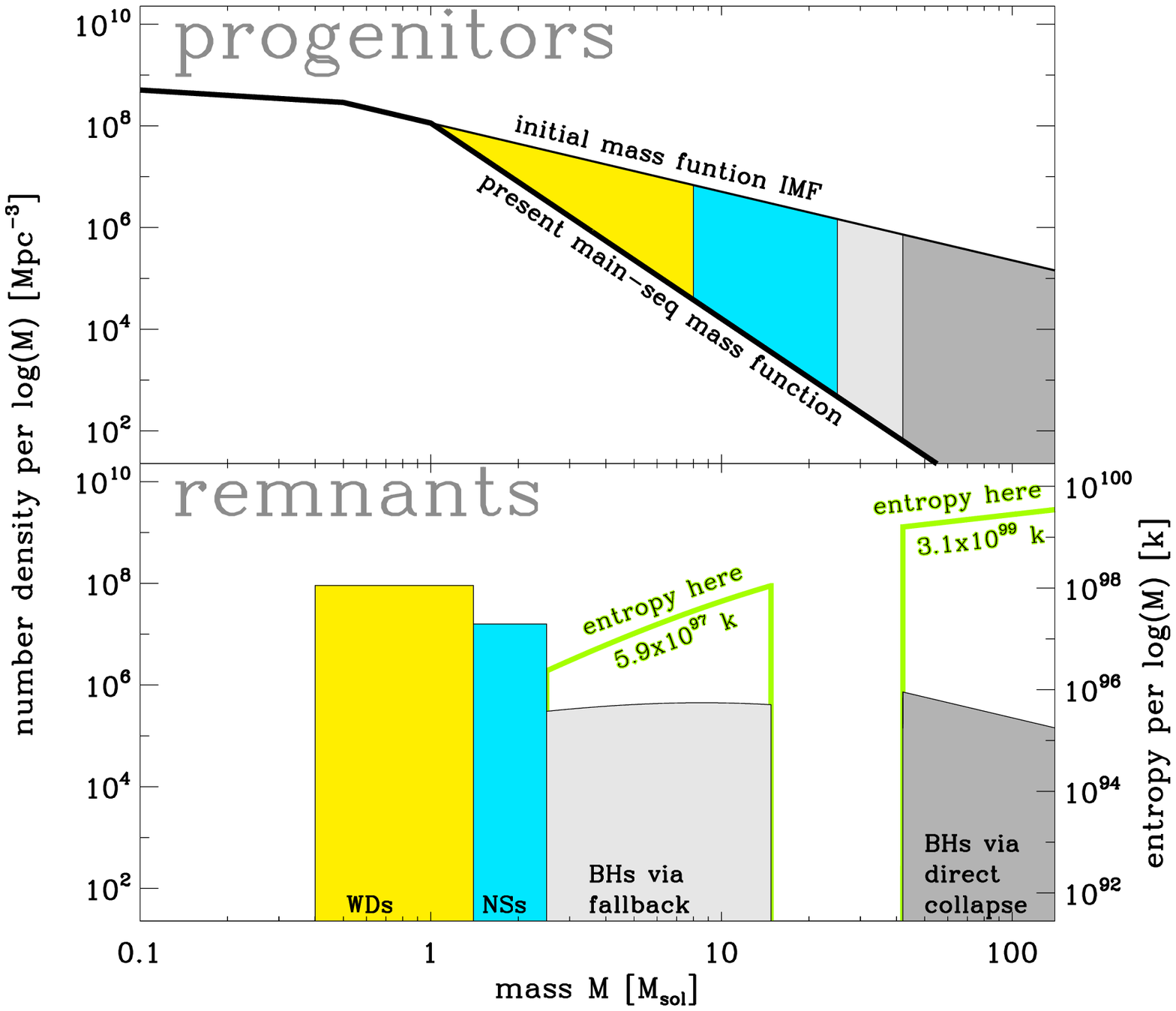}
               		\caption{Progenitors in the IMF (top panel) evolve into the distribution of remnants in 
               		the bottom panel. The shape of the present main-sequence mass function differs from 
               		that of the IMF (top panel) by the stars that have died leaving
               		white dwarfs (yellow), neutron stars (blue), and black holes (light and dark gray). 
               		The present distribution of remnants is shown in the bottom panel. Black 
               		holes in the range $2.5 M_{\odot} \lsim M \lsim 15 M_{\odot}$ (light gray) have been 
               		observationally confirmed. They form from progenitors in the range 
               		$25 M_{\odot} \lsim M \sim 42 M_{\odot}$ via core collapse supernova and fallback, and we
			calculate their entropy to be $5.9\xt{97^{+0.6}_{-1.2}} k$. Progenitors above 
			about $42\ M_{\odot}$ may evolve directly to black holes without significant 
			loss of mass (dark gray) and may carry much more entropy, but this population has 
			not been observed. The green curve, whose axis is on the right, shows 
			the mass distribution of stellar black hole entropies in the observable universe.}
               		\label{fig:sbhentropy}
       		\end{center}
	\end{figure}
}{
	\begin{figure}[!hbtp]
       		\begin{center}
               		\includegraphics[width=\linewidth]{full_imf_2panel_bw.ps}
               		\caption{Progenitors in the IMF (top panel) evolve into the distribution of remnants in 
               		the bottom panel. The shape of the present main-sequence mass function differs from 
               		that of the IMF (top panel) by the stars that have died leaving
               		white dwarfs (dotted), neutron stars (striped), and black holes (light and dark gray). 
               		The present distribution of remnants is shown in the bottom panel. Black 
               		holes in the range $2.5 M_{\odot} \lsim M \lsim 15 M_{\odot}$ (light gray) have been 
               		observationally confirmed. They form from progenitors in the range 
               		$25 M_{\odot} \lsim M \sim 42 M_{\odot}$ via core collapse supernova and fallback, and we 
               		calculate their entropy to be $5.9\xt{97^{+0.6}_{-1.2}} k$. Progenitors above 
               		about $42\ M_{\odot}$ may evolve directly to black holes without significant 
               		loss of mass (dark gray) and may carry much more entropy, but this population has
               		not been observed. The gray curve, whose axis is on the right, shows 
			the mass distribution of stellar black hole entropies in the observable universe. 
			\textit{[See the electronic edition of the journal for a color version of this figure.]}}
               		\label{fig:sbhentropy}
       		\end{center}
	\end{figure}
}

\ifthenelse{\boolean{colver}} { 
The yellow fill }{
The dotted fill }
in the top panel represents stars of mass $1 M_{\odot} \lsim M \lsim 8 M_{\odot}$, 
which died leaving white dwarf remnants of mass $M \lsim 1.4 M_{\odot}$ 
\ifthenelse{\boolean{colver}} { 
(yellow fill, bottom panel). }{
(dotted fill, bottom panel). }
\ifthenelse{\boolean{colver}} { 
The blue fill }{
The striped fill }
represents stars of mass $8 M_{\odot} \lsim M \lsim 25 M_{\odot}$, 
which died and left neutron star remnants of mass $1.4 M_{\odot} \lsim M \lsim 2.5 M_{\odot}$. 
The light gray area represents stars of mass $25 M_{\odot} \lsim M \lsim 42 M_{\odot}$
which became black holes of mass $2.5 M_{\odot} \lsim M \lsim 15 M_{\odot}$ via supernovae 
(here we use the simplistic final-initial mass function of \citet{Fryer2001}). 
Stars larger than $\sim 42 M_{\odot}$ collapse directly to black holes, without supernovae, 
and therefore retain most of their mass (dark gray regions; \citealt{Fryer2001,Heger2005}).

Integrating Equation (\ref{eq:areaent}) over stellar black holes in the range $M \le 15 M_{\odot}$ 
(the light gray fill in the bottom panel of Figure \ref{fig:sbhentropy}) we find
\begin{eqnarray}
	s_{\mathrm{SBH}\ (M<15M_{\odot})} & = & 1.6\xt{17^{+0.6}_{-1.2}}\ k\ m^{-3}, \label{eq:sbhentropy} \\
	S_{\mathrm{SBH}\ (M<15M_{\odot})} & = & 5.9\xt{97^{+0.6}_{-1.2}}\ k,
\end{eqnarray}
which is comparable to previous estimates of the stellar black hole entropy (see 
Table \ref{tab:currententropy}). 
Our uncertainty is dominated by uncertainty in the slope of the IMF, but also includes 
uncertainty in the normalization of the mass functions and uncertainty in the volume of 
the observable universe.

If the IMF extends beyond $M \gsim 42 M_{\odot}$ 
as in Figure \ref{fig:sbhentropy}, then these higher mass black holes (the dark gray
fill in the bottom panel of Figure \ref{fig:sbhentropy}) may contain more
entropy than black holes of mass $M < 15\ M_{\odot}$ (Equation \ref{eq:sbhentropy}). For example, if the 
Salpeter IMF is reliable to $M=140\ M_{\odot}$ (the Eddington limit and the edge of Figure
\ref{fig:sbhentropy}), then black holes in the mass range $42$ - $140\ M_{\odot}$ would 
contribute about $3.1\xt{99^{+0.8}_{1.6}}\ k$ to the entropy of the observable universe. 
Significantly less is known about this potential population, 
and should be considered a tentative contribution in Table \ref{tab:currententropy}.

\subsection{Supermassive Black Holes} \label{smbh}

Previous estimates of the SMBH entropy \citep{Penrose2004,Frampton2008,Frampton2008b} 
have assumed a typical SMBH mass and a number density and yield 
$S_{\mathrm{SMBH}} = 10^{101}-10^{103} k$. Below we use the SMBH mass function as measured 
recently by \citet{Graham2007}. 
Assuming a three-parameter Schechter function
\begin{eqnarray}
	\frac{d n}{d \log(M)} = \phi_* \left( \frac{M}{M_*} \right)^{\alpha + 1} \exp{\left[ 1 - \left(\frac{M}{M_*}\right) \right]}
\end{eqnarray} 
(number density per logarithmic mass interval) they find 
$\phi_* = 0.0016 \pm 0.0004\ {Mpc}^{-3}$, 
$M_* = 2.9 \pm 0.7 \xt{8}\ M_{\odot}$, and 
$\alpha = -0.30 \pm 0.04$. 
The data and best-fit model are shown in black in Figure \ref{fig:bhentropy}. 

\ifthenelse{\boolean{colver}} {
	\begin{figure}[!hbtp]
       		\begin{center}
               		\includegraphics[width=\linewidth]{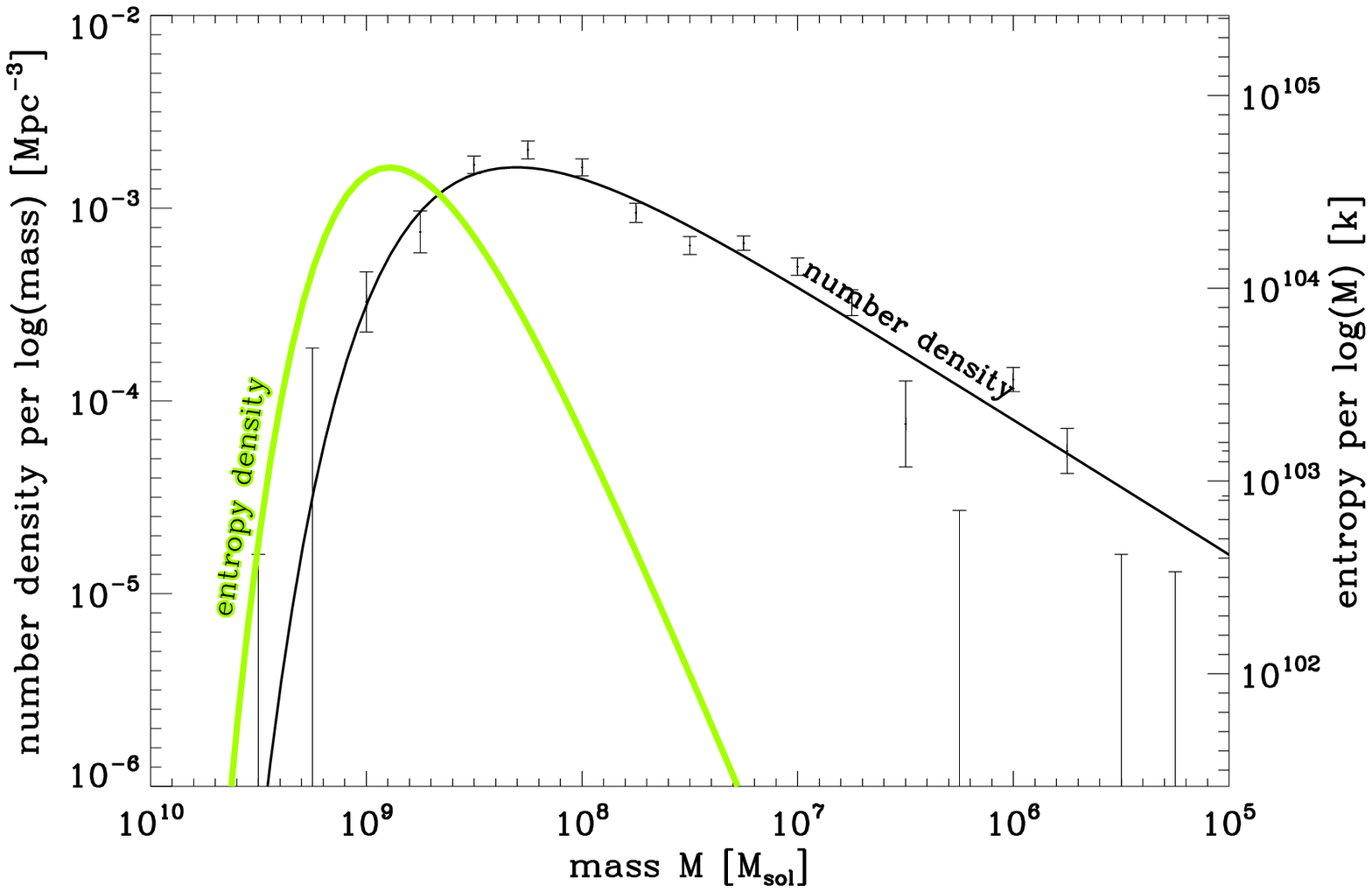}
               		\caption{The black curve,
               		whose axis is on the left, is the SMBH mass function from \citet{Graham2007}, i.e.,
               		the number of supermassive black holes per ${Mpc}^{3}$ per logarithmic mass interval. 
               		The green curve, whose axis is on the right, shows the mass distribution 
			of SMBH entropies in the observable universe.}
               		\label{fig:bhentropy}
       		\end{center}
	\end{figure}
}{
	\begin{figure}[!hbtp]
       		\begin{center}
               		\includegraphics[width=\linewidth]{SMBH_mass_function_bw.ps}
               		\caption{The black curve,
               		whose axis is on the left, is the SMBH mass function from \citet{Graham2007}, i.e.,
               		the number of supermassive black holes per ${Mpc}^{3}$ per logarithmic mass interval. 
               		The gray curve, whose axis is on the right, shows the mass distribution 
			of SMBH entropies in the observable universe. \textit{[See the electronic 
			edition of the journal for a version of this figure.]}}
               		\label{fig:bhentropy}
       		\end{center}
	\end{figure}
}

We calculate the SMBH entropy density by integrating Equation (\ref{eq:areaent}) over
the SMBH mass function,
\begin{eqnarray}
	s 	
		& = & \frac{4 \pi k G}{c \hbar}\ \int M^2 \left(\frac{d n}{d \log(M)}\right)\ d\log(M).
\end{eqnarray}
The integrand is plotted using 
\ifthenelse{\boolean{colver}} { 
a green line }{
a gray line }
in Figure \ref{fig:bhentropy} showing that
the contributions to SMBH entropy are primarily due to black holes around 
$\sim 10^9 M_{\odot}$. 
The SMBH entropy is found to be 
\begin{eqnarray}
	s_{\mathrm{SMBH}} & = & 8.4^{+8.2}_{-4.7} \xt{23}\ k\ m^{-3}, \\
	S_{\mathrm{SMBH}} & = & 3.1^{+3.0}_{-1.7} \xt{104}\ k.
\end{eqnarray}
The uncertainty here includes uncertainties in the SMBH mass function and uncertainties
in the volume of the observable universe.
This is at least an order of magnitude larger than previous estimates (see Table \ref{tab:currententropy}). 
The reason for the difference is that the \citep{Graham2007} SMBH mass function contains 
larger black holes than assumed in previous estimates.

\ifthenelse{\boolean{colver}} {
	\begin{figure}[!hbtp]
       		\begin{center}
               		\includegraphics[width=\linewidth]{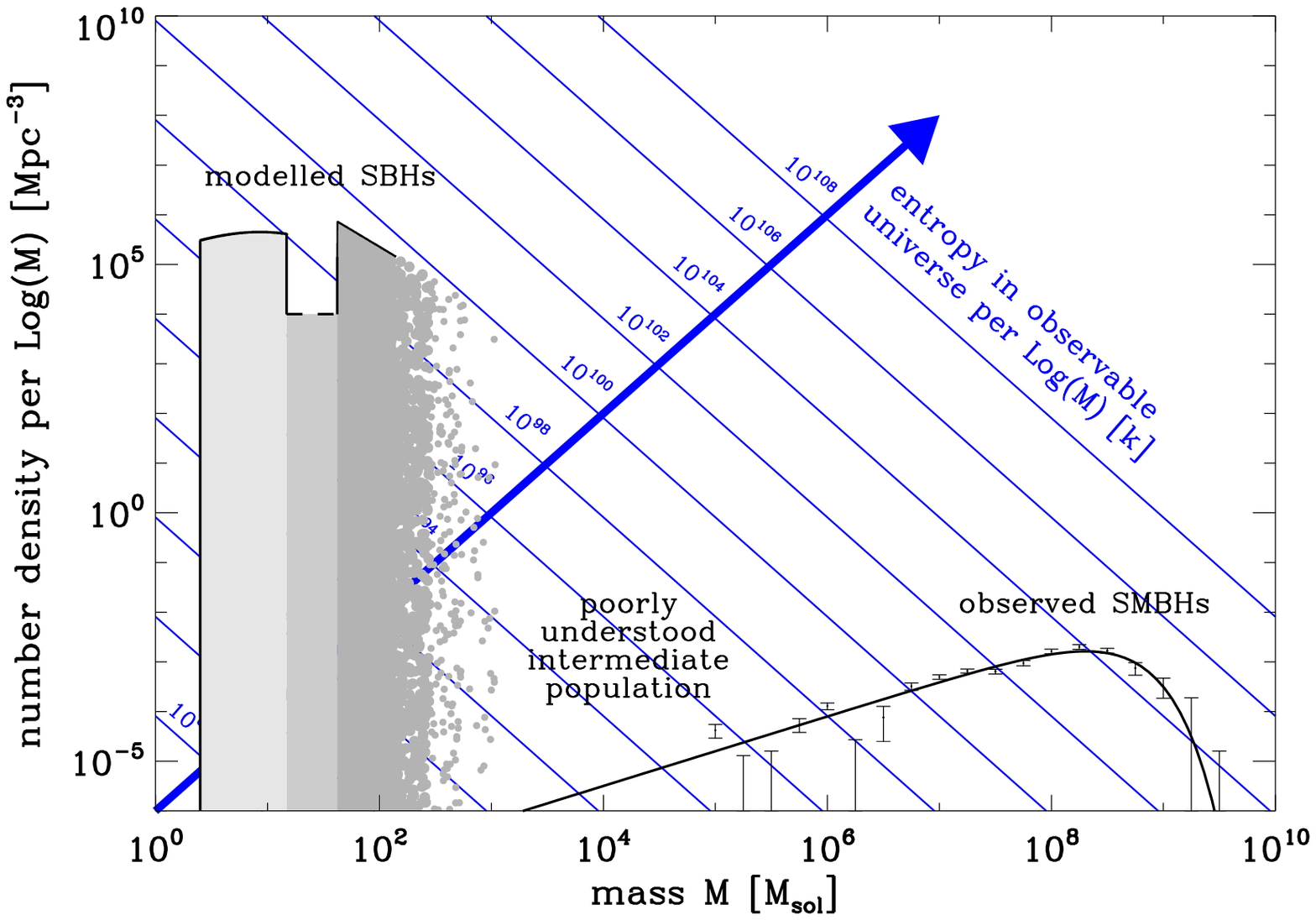}
               		\caption{Whether or not the total black hole entropy is dominated by SMBHs
               		depends on the yet-unquantified number of intermediate mass black holes.}
               		\label{fig:combinedbhentropy}
       		\end{center}
	\end{figure}
}{
	\begin{figure}[!hbtp]
       		\begin{center}
               		\includegraphics[width=\linewidth]{combination_fried_rice_bw.ps}
               		\caption{Whether or not the total black hole entropy is dominated by SMBHs
               		depends on the yet-unquantified number of intermediate mass black holes. 
			\textit{[See the electronic edition of the journal for a color version of this figure.]}}
               		\label{fig:combinedbhentropy}
       		\end{center}
	\end{figure}
}

\citet{Frampton2009,Frampton2009b} has suggested that intermediate mass black holes 
in galactic halos may contain more entropy than SMBHs in galactic cores. For example, according 
to the massive astrophysical compact halo object (MACHO) explanation of dark matter, intermediate 
mass black holes in the mass range $10^2$ - $10^5 \ M_{\odot}$ may constitute 
dark matter. Assuming $10^5 \ M_{\odot}$ black holes, these objects would contribute up to 
$10^{106}\ k$ to the entropy of the observable universe \citep{Frampton2009b}.
Whether or not this is so depends on the number density and mass distribution of this population. 
Figure \ref{fig:combinedbhentropy} combines Figures \ref{fig:sbhentropy} and \ref{fig:bhentropy} and 
shows what intermediate black hole number densities would be required.

\section{The Entropy of the Cosmic Event Horizon and its Interior} \label{scheme2} \label{ceh}

In this section we calculate the entropy budget for scheme 2 (refer  to discussion in Section \ref{frwgsl}). 
Scheme 2 differs from scheme 1 in two ways: first, along with the components previously considered 
(and listed in Table \ref{tab:currententropy}), here we consider the CEH as an 
additional entropy component; and second, the volume of interest is that within the event horizon 
not the particle horizon (or observable universe).

The proper distance to the CEH is generally time-dependent, increasing when the 
universe is dominated by an energy component with an equation of state $w>-1$ (radiation 
and matter) and remaining constant when the universe is dark energy dominated (assuming 
a cosmological constant, $w=-1$). Since our universe is presently entering dark energy domination, 
the growth of the event horizon has 
slowed, and it is almost as large now as it will ever become (bottom panel of Figure \ref{fig:FRW}). 
In the Appendix, we calculate the present radius and volume of the CEH
\begin{eqnarray}
	R_{\mathrm{CEH}} = 15.7 \pm 0.4\ {Glyr},
\end{eqnarray}
\begin{eqnarray}
	V_{\mathrm{CEH}} & = & 1.62 \pm 0.12 \xt{4}\ {Glyr}^3 \nonumber \\
		& = & 1.37 \pm 0.10 \xt{79}\ m^3.
\end{eqnarray}
We also calculate the present entropy of the CEH (following \citealt{Gibbons1977}),
\begin{eqnarray}
	S_{\mathrm{CEH}} & = & \frac{k c^3}{G \hbar} \frac{A}{4} \nonumber \\
		& = & \frac{k c^3}{G \hbar} \pi R_{\mathrm{CEH}}^2 \\
		& = & 2.6 \pm 0.3 \xt{122}\ k. \nonumber
\end{eqnarray}

Entropies of the various components within the CEH are calculated using the 
entropy densities $s_{i}$ from Section \ref{scheme1}: 
\begin{eqnarray}
	S_{i} & = & s_{i} V_{\mathrm{CEH}}
\end{eqnarray}
Table \ref{tab:scheme2budget} shows that the cosmic event horizon contributes almost $20$
orders of magnitude more entropy than the next largest contributor, supermassive black holes.

\begin{deluxetable*}{l l}
\tablecaption{Entropy of the Event Horizon and the Matter Within it (Scheme 2 Entropy Budget) \label{tab:scheme2budget}}
\tablehead{
	\colhead{Component} &
	\colhead{Entropy $S$ $[k]$}
}
\startdata
	Cosmic Event Horizon
		& $2.6\pm0.3\xt{122}$
		\\
	SMBHs
		& $1.2^{+1.1}_{-0.7} \xt{103}$
		\\
	Stellar BHs $(2.5-15\ M_{\odot})$
		& $2.2\xt{96^{+0.6}_{-1.2}}$		
		\\
	Photons
		& $2.03\pm0.15\xt{88}$				
		\\
	Relic Neutrinos
		& $1.93\pm0.15\xt{88}$
		\\
	\addthis{WIMP} Dark Matter
		& $6\xt{86\pm1}$	
		\\
	Relic Gravitons
		& $2.3\xt{86^{+0.2}_{-3.1}}$				
		\\
	ISM and IGM
		& $2.7 \pm 2.1 \xt{80}$	
		\\
	Stars
		& $3.5 \pm 1.7 \xt{78}$
		\\
	\hline
	{\bf Total}
		& \boldmath$2.6\pm0.3\xt{122}$
		\\
	\hline
	\addthis{Tentative Components:}
		&
		\\
	\addthis{Massive Halo BHs $(10^5\ M_{\odot})$}
		& \addthis{$10^{104}$}
		\\
	\addthis{Stellar BHs $(42-140\ M_{\odot})$}
		& \addthis{$1.2\xt{98^{+0.8}_{-1.6}}$}
\enddata
\tablecomments{
	This budget is dominated by the 
	cosmic event horizon entropy. 
	While the CEH entropy should be considered as an additional component in
	scheme 2, it also corresponds to the holographic bound \citep{tHooft1993} on the 
	possible entropy of the other components and may represent a significant overestimate.
	\removethis{Stellar black holes in the mass range $42-140\ M_{\odot}$ (marked with an $*$) are 
	included tentatively since their existence is speculative.} \addthis{Massive halo black holes at 
	$10^5\ M_{\odot}$ and stellar black holes in the range $42-140\ M_{\odot}$ are included 
	tentatively since their existence is speculative.}
	}
\end{deluxetable*}

\section{Discussion} \label{discussion}

The second law of thermodynamics holds that the entropy of an isolated system
increases or remains constant, but does not decrease.
This has been applied to the large-scale universe in at least two ways (Equation
\ref{eq:one} and \ref{eq:two}). The first scheme requires the entropy in a comoving 
volume of the universe to not decrease. The second scheme requires the entropy 
of matter contained within the event horizon, plus the entropy of the event horizon, 
to not decrease.

We have calculated improved estimates of the current entropy budget under scheme 1
(normalized to the current observable universe) and scheme 2. These are given in Tables 
\ref{tab:currententropy} and \ref{tab:scheme2budget}, respectively.




The entropy of dark matter has not been calculated previously. 
We find that dark matter contributes $10^{88\pm1}\ k$ to the entropy of the 
observable universe. 
We note that the neutrino and dark matter estimates do not include an increase 
due to their infall into gravitational potentials during structure formation. It is not
clear to us a priori whether this non-inclusion is significant, but it may be since 
both components are presently non-relativistic. 
This should be investigated in future work.

Previous estimates of the relic graviton entropy have assumed that only the
known particles participate in the relativistic fluid of the early universe at $t \gsim t_{\mathrm{planck}}$. 
In terms of the number of relativistic degrees of freedom, this means 
$g_{*S} \rightarrow 106.75$ at high temperatures. However, additional particles are 
expected to exist, and thus $g_{*S}$ is expected to become larger as 
$t \rightarrow t_{\mathrm{planck}}$. In the present work, we have calculated the relic graviton 
entropy corresponding to three high-energy extrapolations of $g_{*S}$ (constant, linear 
growth and exponential growth) and reported the corresponding graviton temperatures 
and entropies. 


In this paper, we have computed the entropy budget of the observable universe 
today $S_{\mathrm{obs}}(t=t_0)$. Figure \ref{fig:eoftimes1} illustrates the evolution of the entropy 
budget under scheme 1, i.e., the entropy in a comoving volume (normalized to the current 
observable universe). For simplicity, we have included only the most important 
components.
\ifthenelse{\boolean{colver}} {
	\begin{figure}[!hbtp]
       		\begin{center}
               		\includegraphics[width=\linewidth]{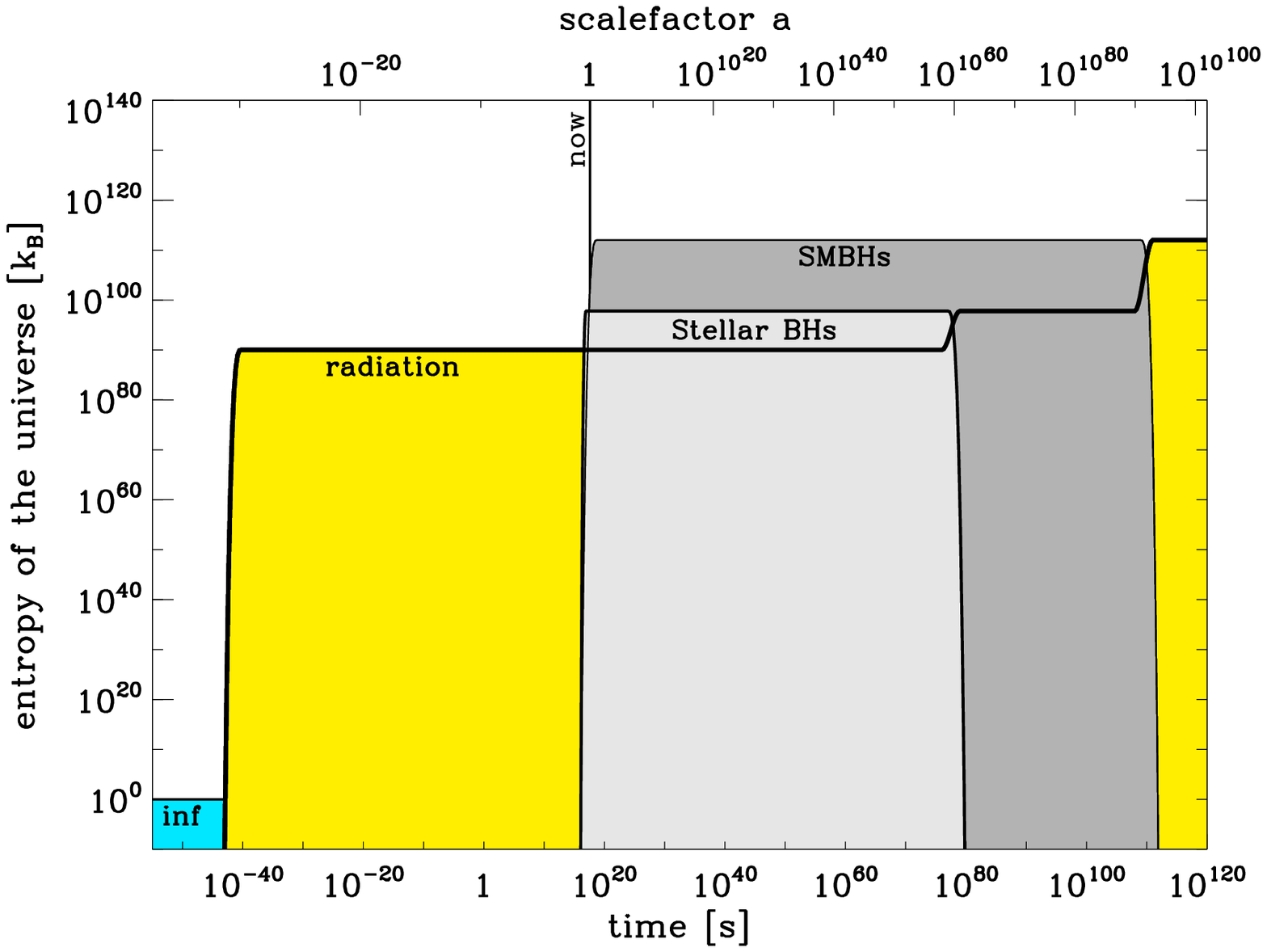}
               		\caption{The entropy in a comoving volume (normalized to the 
               		present observable universe). This figure illustrates the time-dependence
               		of the scheme 1 entropy budget. N.B.\ $10^{10^{100}} = 1$ googolplex.}
               		\label{fig:eoftimes1}
       		\end{center}
	\end{figure}
}{
	\begin{figure}[!hbtp]
       		\begin{center}
               		\includegraphics[width=\linewidth]{eoftimes1_bw.ps}
               		\caption{The entropy in a comoving volume (normalized to the 
               		present observable universe). This figure illustrates the time-dependence
               		of the scheme 1 entropy budget. N.B.\ $10^{10^{100}} = 1$ googolplex. 
			\textit{[See the electronic edition of the journal for a color version of this figure.]}}
               		\label{fig:eoftimes1}
       		\end{center}
	\end{figure}
}

At the far-left of the figure, we show a brief period of inflation. During this period
all of the energy is in the inflaton \citep{Guth1981,Linde1982}, which 
has very few degrees of freedom and low entropy 
\ifthenelse{\boolean{colver}} { 
(blue fill; \citealt{Linde2009,Steinhardt2009}).}{
(crosshatched fill; \citealt{Linde2009,Steinhardt2009}).}
Inflation ends 
with a period of reheating somewhere between the Planck scale ($10^{-45}s$) 
and the GUT scale ($10^{-35}s$), during which the inflaton's energy is transferred
into a relativistic fluid 
\ifthenelse{\boolean{colver}} { 
(yellow fill). }{
(dotted fill). }
During reheating, the entropy increases by
many orders of magnitude. After reheating, the constitution of the relativistic 
fluid continues to change, but the changes occur reversibly and do not increase 
the entropy.

After a few hundred million years ($\sim 10^{16}s$), the first stars form from collapsing 
clouds of neutral hydrogen and helium. Shortly thereafter the first black holes form. 
The entropy in stellar black holes (light gray) and SMBHs (dark
gray) increases rapidly during galactic evolution. The budget given in Table 
\ref{tab:currententropy} is a snapshot of the entropies at the present time ($4.3\xt{17}s$). 
Over the next $10^{26}s$, the growth of structures 
larger than about $10^{14}\ M_{\odot}$ will be halted by the acceleration of the 
universe. Galaxies within superclusters will merge and those in the outer limits 
will be ejected. The final masses of SMBHs will 
be $\sim 10^{10} M_{\odot}$ \citep{Adams1997} with the entropy dominated by
those with $M \sim 10^{12} M_{\odot}$.

Stellar black holes will evaporate away into Hawking radiation in about $10^{80}s$
and SMBHs will follow in $10^{110}s$. The decrease in
black hole entropy is accompanied by a compensating increase in radiation entropy.
The thick black line in Figure \ref{fig:eoftimes1} represents the radiation entropy growing 
as black holes evaporate.
The asymptotic future of the entropy budget, under scheme 1, will be radiation 
dominated.

Figure \ref{fig:eoftimes2} illustrates the evolution of the entropy budget under 
scheme 2, i.e., the entropy within the CEH, plus the 
entropy of the CEH. 
\ifthenelse{\boolean{colver}} {
	\begin{figure}[!hbtp]
       		\begin{center}
               		\includegraphics[width=\linewidth]{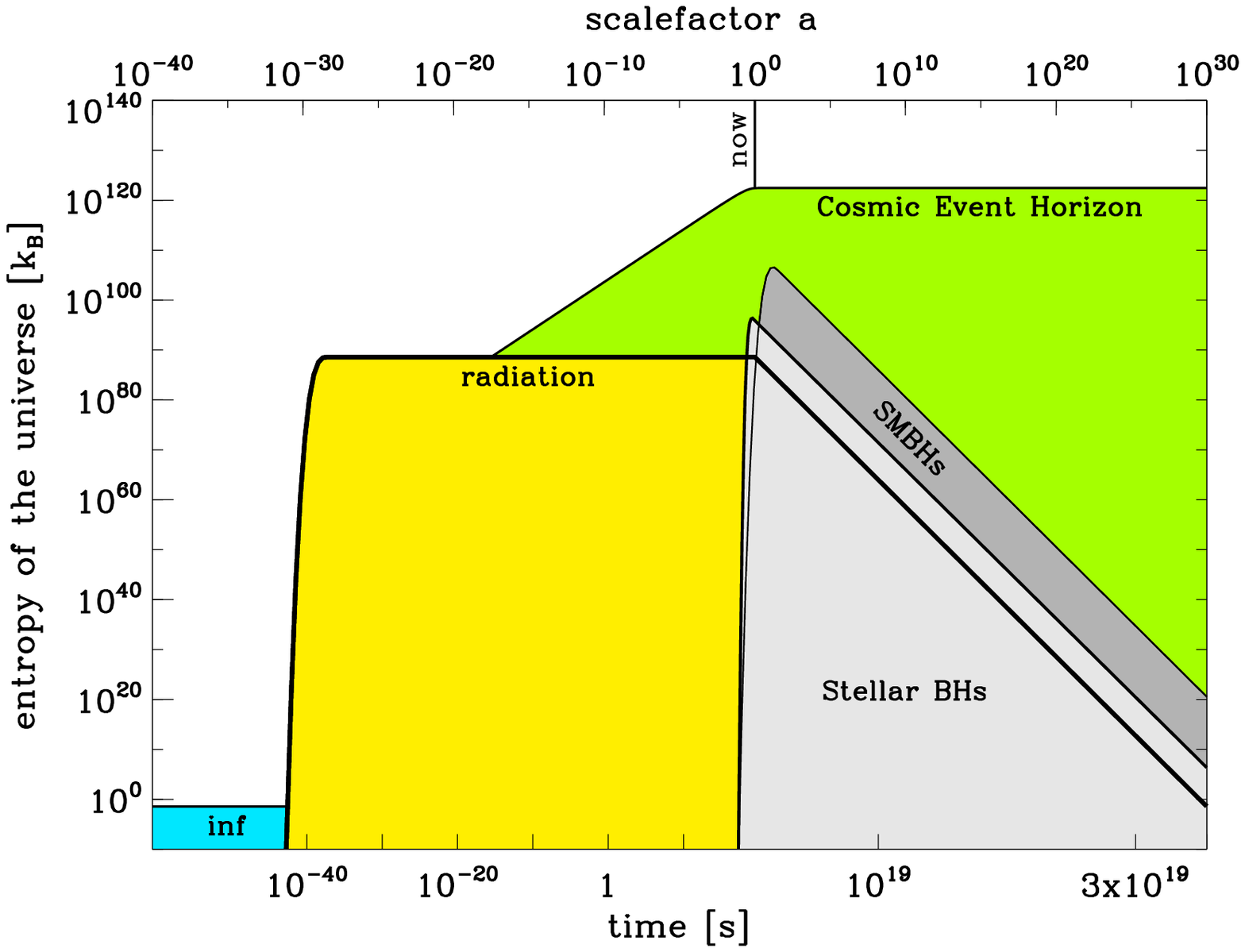}
               		\caption{Entropy of matter within the CEH, and the
               		entropy of the cosmic event horizon. This figure illustrates the time
               		dependence of the scheme 2 entropy budget. Note: the horizontal 
               		axis is shorter than in Figure \ref{fig:eoftimes1}.}
               		\label{fig:eoftimes2}
       		\end{center}
	\end{figure}
}{
	\begin{figure}[!hbtp]
       		\begin{center}
               		\includegraphics[width=\linewidth]{eoftimes2_bw.ps}
               		\caption{Entropy of matter within the CEH, and the
               		entropy of the CEH. This figure illustrates the time
               		dependence of the scheme 2 entropy budget. Note: the horizontal 
               		axis is shorter than in Figure \ref{fig:eoftimes1}. \textit{[See the electronic 
			edition of the journal for a color version of this figure.]}}
               		\label{fig:eoftimes2}
       		\end{center}
	\end{figure}
}

Whereas in scheme 1 we integrate over a constant comoving volume, here
the relevant volume is the event horizon. The event horizon is discussed in some 
detail in the Appendix. During radiation domination, the comoving 
radius of the CEH is approximately constant (the proper 
distance grows as $R_{\mathrm{CEH}} \propto a$) and in the dark energy dominated 
future, it is a constant proper distance ($R_{\mathrm{CEH}} = \mathrm{constant}$). 
The few logarithmic decades around the present time cannot be described
well by either of these.

Since the event horizon has been approximately comoving in the past, the 
left half of Figure \ref{fig:eoftimes2} is almost the same as in Figure \ref{fig:eoftimes1}
except that we have included the event horizon entropy 
\ifthenelse{\boolean{colver}} { 
(green fill). }{
(striped fill). } 
The event horizon entropy dominates this budget from about $10^{-16}s$.

After dark energy domination sets in, the CEH becomes a
constant proper distance. The expansion of the universe causes comoving 
objects to recede beyond the CEH. On average, the number 
of galaxies, black holes, photons etc.\ within our CEH decreases 
as $a^{-3}$. The stellar and SMBH entropy contained
within the CEH decreases accordingly (decreasing gray filled regions).

The decreasing black hole entropy (as well as other components not shown) 
is compensated by the asymptotically growing CEH entropy 
(demonstrated explicitly for a range of scenarios in \citealt{Davis2003}), and thus 
the second law of thermodynamics is satisfied. 
See \citet[in preparation]{EganLineweaver2009b} for further discussion of the time-dependence 
of the entropy of the universe.

\section*{Acknowledgments}

We are grateful for many useful discussions with Tamara Davis, 
Ken Freeman, \addthis{Pat Scott,} Geoff Bicknell, Mike Turner, Andrei Linde, and Paul 
Steinhardt. C.A.E. thanks Anna Fransson for financial support and the Research School of 
Astronomy and Astrophysics, Australian National University, for its hospitality during the 
preparation of this paper.

\section*{Appendix: The observable universe and the cosmic event horizon} \label{phceh}

Here we calculate the radius and volume of the observable universe (for use in Section \ref{scheme1});
and we calculate the radius, volume, and entropy of the CEH (for use in Section \ref{scheme2}).
We use numerical methods to track the propagation of errors from the cosmological parameters. 


The radius of the observable universe (or particle horizon) is
\begin{eqnarray}
	R_{\mathrm{obs}} = a(t) \int_{t'=0}^{t} \frac{c}{a(t')} dt'. \label{eq:robs}
\end{eqnarray}
Here $a(t)$ is the time-dependent scalefactor of the universe given by the Friedmann equation for a 
flat cosmology
\begin{eqnarray}
	\frac{da}{dt} = \sqrt{\frac{\Omega_{r}}{a^2} + \frac{\Omega_{m}}{a} + \frac{\Omega_{\Lambda}}{a^{-2}}}.
\end{eqnarray}
Hubble's constant and the 
matter density parameter are taken from \citet{Seljak2006}: 
$h = H / 100\ km\ s^{-1}\ Mpc^{-1} = 0.705 \pm 0.013$, $\omega_m = \Omega_m h^2 = 0.136 \pm 0.003$.
The radiation density is calculated from the observed CMB temperature, 
$T_{\mathrm{CMB}} = 2.725 \pm 0.002\ K$ \citep{Mather1999}, using
$\Omega_{r} = \frac{8 \pi G}{3 H^2} \frac{\pi^2 k^4 T^4}{15 c^5 \hbar^3}$.
The vacuum energy density parameter is determined by flatness, 
$\Omega_{\Lambda} = 1-\Omega_{r}-\Omega_{m}$.

A distribution of $R_{\mathrm{obs}}$ values is built up by repeatedly evaluating Equation (\ref{eq:robs})
at the present time (defined by $a(t_0)=1$) using 
cosmological parameters randomly selected from the allowed region of $h - \omega_{m} - T_{\mathrm{CMB}}$ 
parameter space (assuming uncorrelated Gaussian errors in these parameters). We find
\begin{eqnarray}
	R_{\mathrm{obs}} = 46.9 \pm 0.4\ {Glyr}
\end{eqnarray}
with an approximately Gaussian distribution. The quoted confidence interval here, and elsewhere
in this Appendix, is $1\sigma$.
The volume of the observable universe $V_{\mathrm{obs}}$ is calculated using the normal formula for 
the volume of a sphere. 
\begin{eqnarray}
	V_{\mathrm{obs}} & = & 43.2 \pm 1.2 \xt{4}\ {Glyr}^3 \nonumber \\ 
		& = & 3.65 \pm 0.10 \xt{80}\ m^3 
\end{eqnarray}
See Figure \ref{fig:vphveh}.
Uncertainty in $R_{\mathrm{obs}}$ and $V_{\mathrm{obs}}$ is predominantly due to uncertainty in $\omega_m$ 
however $h$ also makes a non-negligible contribution.

\ifthenelse{\boolean{colver}} {
	\begin{figure}[!hbtp]
       		\begin{center}
               		\includegraphics[width=\linewidth]{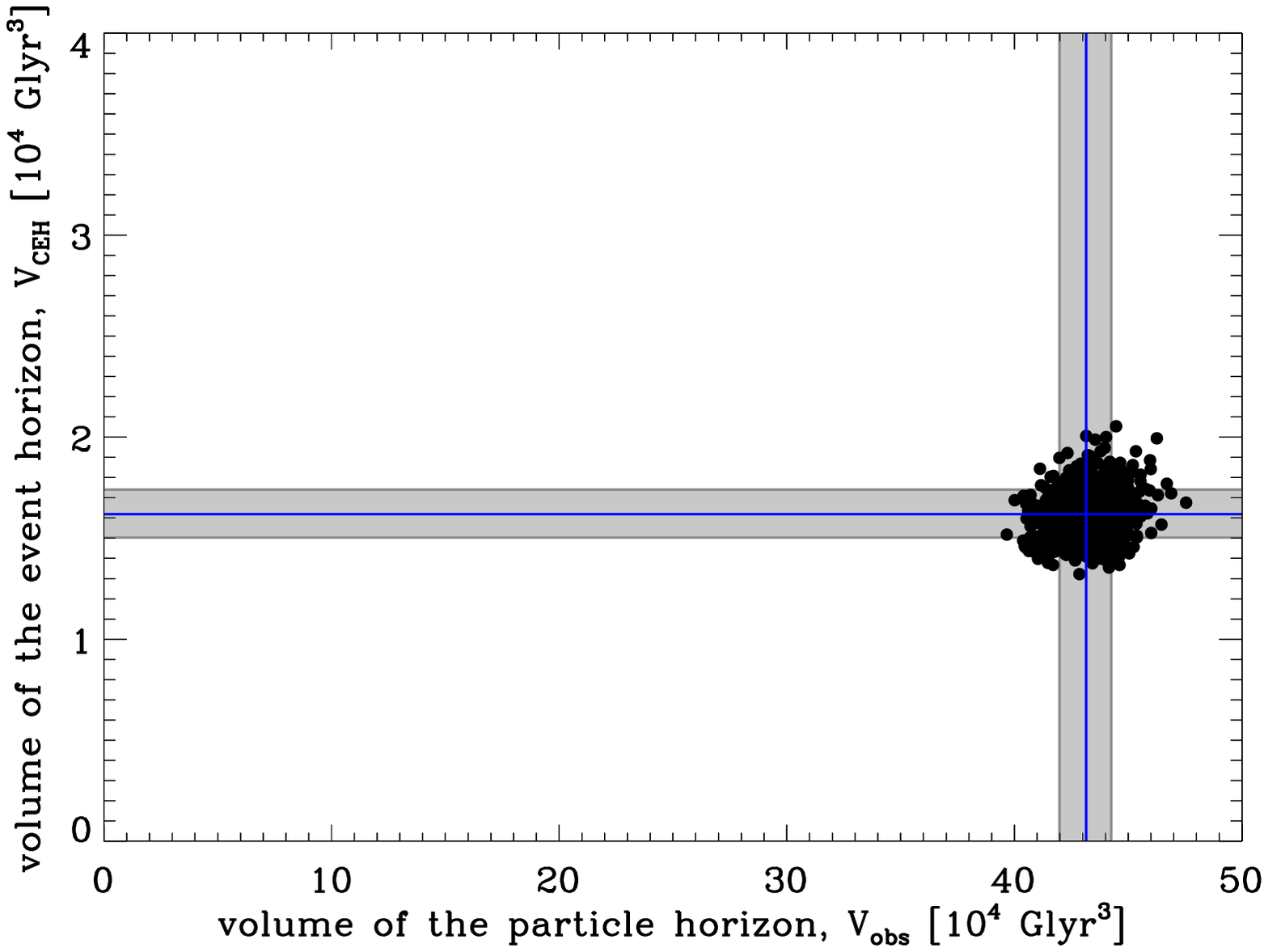}
               		\caption{Eight hundred realizations of $V_{\mathrm{obs}}$ and $V_{\mathrm{CEH}}$ indicate the volume of the 
               		observable universe is $43.2 \pm 1.2 \xt{4}\ {Glyr}^3$ (horizontal axis) and the volume of 
               		the cosmic event horizon is $V_{\mathrm{CEH}}=1.62 \pm 0.12 \xt{4}\ {Glyr}^3$ (vertical axis). We 
               		note that there is only a weak correlation between uncertainties in the two volumes.}
               		\label{fig:vphveh}
       		\end{center}
	\end{figure}
}{
	\begin{figure}[!hbtp]
       		\begin{center}
               		\includegraphics[width=\linewidth]{vph_and_veh_bw.ps}
               		\caption{Eight hundred realizations of $V_{\mathrm{obs}}$ and $V_{\mathrm{CEH}}$ indicate the volume of the 
               		observable universe is $43.2 \pm 1.2 \xt{4}\ {Glyr}^3$ (horizontal axis) and the volume of 
               		the cosmic event horizon is $V_{\mathrm{CEH}}=1.62 \pm 0.12 \xt{4}\ {Glyr}^3$ (vertical axis). We 
               		note that there is only a weak correlation between uncertainties in the two volumes. 
			\textit{[See the electronic edition of the journal for a color version of this figure.]}}
               		\label{fig:vphveh}
       		\end{center}
	\end{figure}
}


The radius of the CEH at time $t$ is given by integrating along a photon's 
world line from the time $t$ to the infinite future.
\begin{eqnarray} \label{eq:deh}
	R_{\mathrm{CEH}} = a(t_{now}) \int_{t=t_{now}}^{\infty} \frac{c}{a(t)} dt
\end{eqnarray}
This integral is finite because the future of the universe is dark energy dominated.
Using the same methods as for the observable universe, we find the present radius 
and volume of the CEH to be
\begin{eqnarray}
	R_{\mathrm{CEH}} = 15.7 \pm 0.4\ {Glyr},
\end{eqnarray}
and
\begin{eqnarray}
	V_{\mathrm{CEH}} & = & 1.62 \pm 0.12 \xt{4}\ {Glyr}^3, \nonumber \\ 
		& = & 1.37 \pm 0.10 \xt{79}\ m^3. 
\end{eqnarray}

\ifthenelse{\boolean{colver}} {
	\begin{figure}[!hbtp]
       		\begin{center}
               		\includegraphics[width=\linewidth]{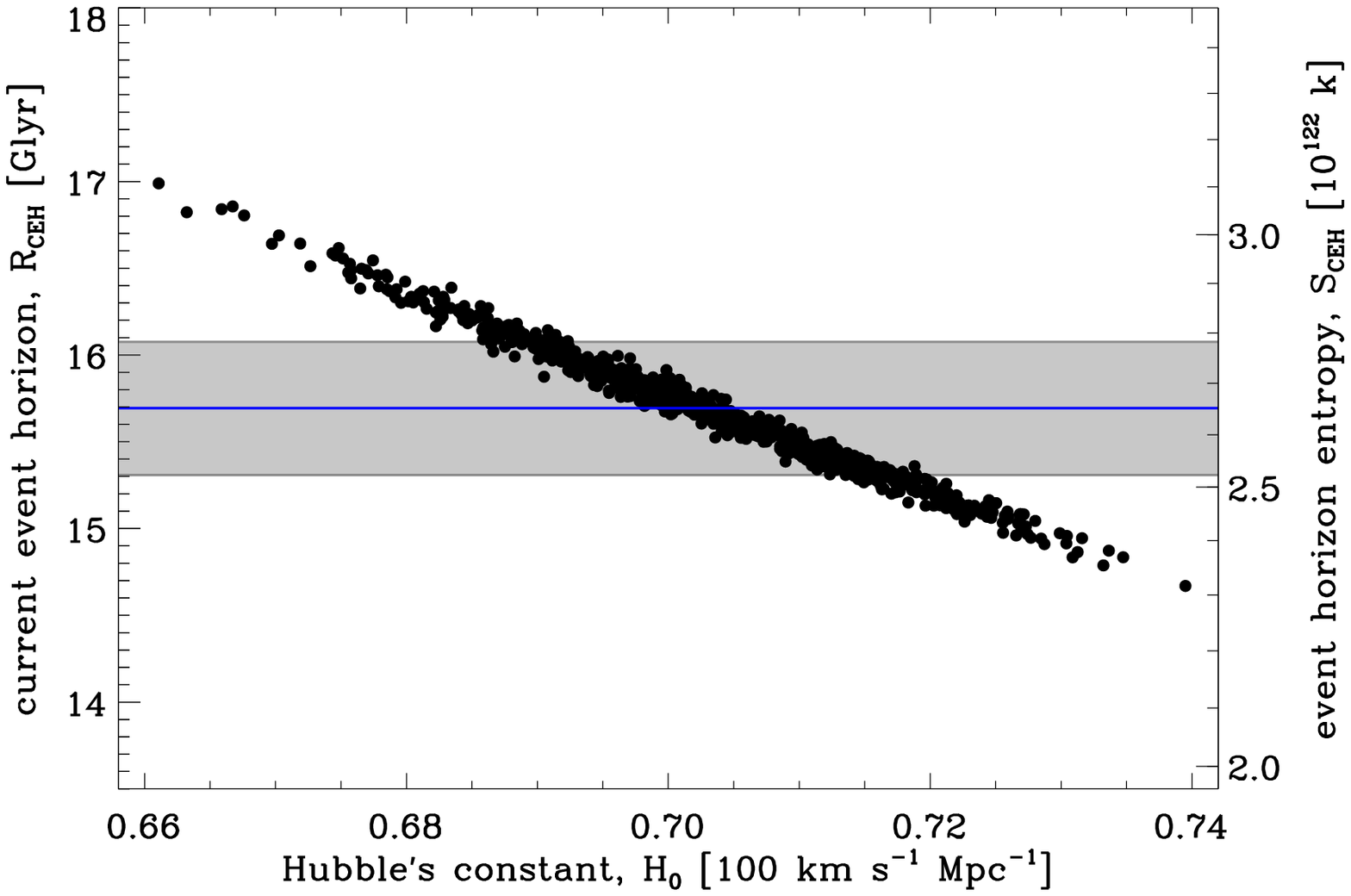}
               		\caption{We find $S_{\mathrm{CEH}} = 2.6 \pm 0.3 \xt{122}\ k$, in agreement with previous 
               		estimates $S_{\mathrm{CEH}} \sim 10^{122}\ k$ \citep{Bousso2007}. Uncertainties in $S_{\mathrm{CEH}}$ 
               		come from uncertainties in $R_{\mathrm{CEH}}$, which are almost exclusively due to uncertainties 
			in $h$.}
               		\label{fig:dehh0}
       		\end{center}
	\end{figure}
}{
	\begin{figure}[!hbtp]
       		\begin{center}
               		\includegraphics[width=\linewidth]{deh_h0_bw.ps}
               		\caption{We find $S_{\mathrm{CEH}} = 2.6 \pm 0.3 \xt{122}\ k$, in agreement with previous 
               		estimates $S_{\mathrm{CEH}} \sim 10^{122}\ k$ \citep{Bousso2007}. Uncertainties in $S_{\mathrm{CEH}}$ 
               		come from uncertainties in $R_{\mathrm{CEH}}$, which are almost exclusively due to uncertainties 
			in $h$. \textit{[See the electronic edition of the journal for a color version of this figure.]}}
               		\label{fig:dehh0}
       		\end{center}
	\end{figure}
}

\ifthenelse{\boolean{colver}} {
	\begin{figure}[!hbtp]
       		\begin{center}
               		\includegraphics[width=\linewidth]{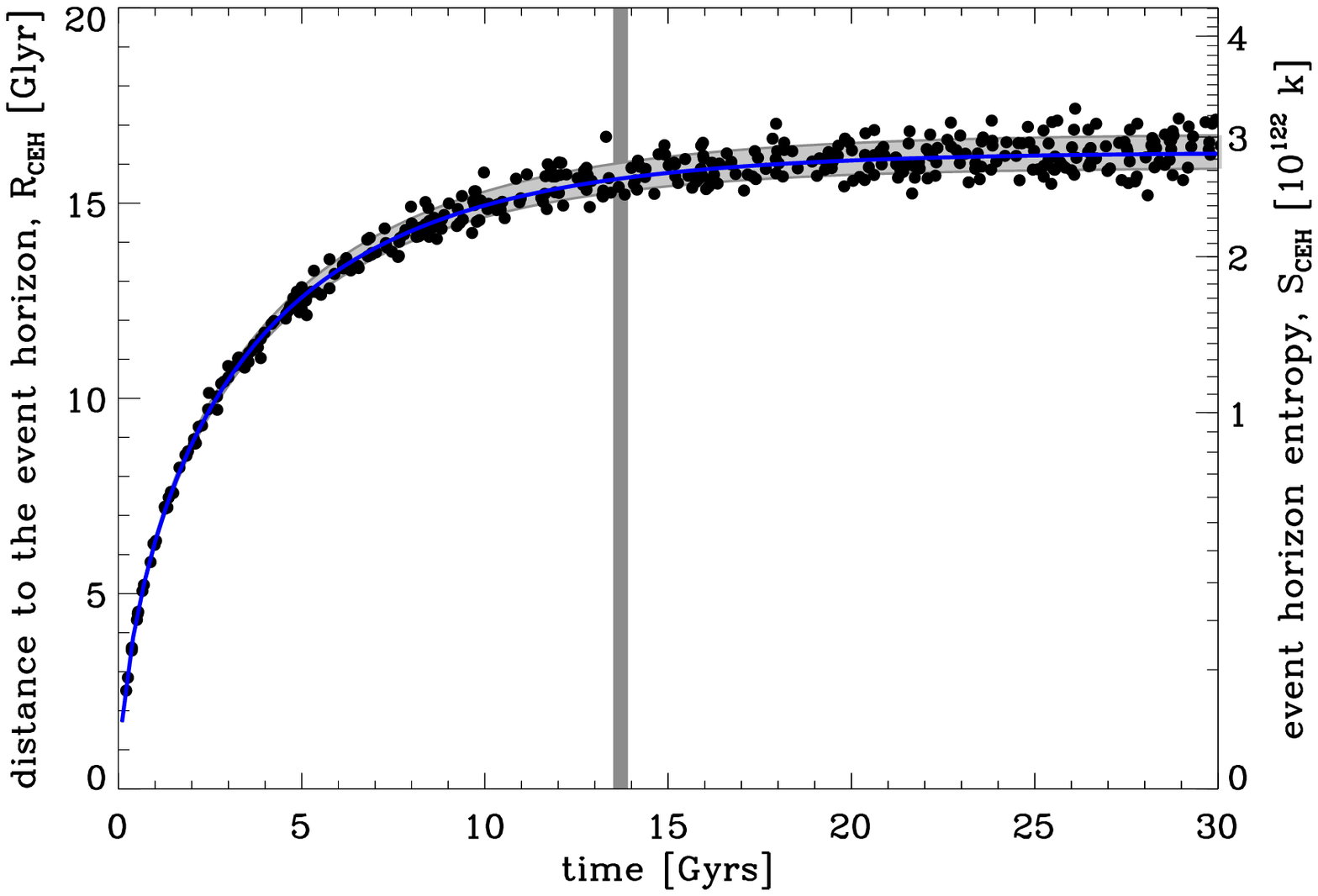}
               		\caption{Proper distance to the event horizon is shown as a function
               		of time. The vertical gray line represents the present age of the universe (and its width, the 
               		uncertainty in the present age).
               		During dark energy domination, the proper radius, proper volume, and entropy 
               		of the CEH will monotonically increase, asymptoting to a constant.
			} \label{fig:deht}
       		\end{center}
	\end{figure}
}{
	\begin{figure}[!hbtp]
       		\begin{center}
               		\includegraphics[width=\linewidth]{deht_bw.ps}
               		\caption{Proper distance to the event horizon is shown as a function
               		of time. The vertical gray line represents the present age of the universe (and its width, the 
               		uncertainty in the present age).
               		During dark energy domination, the proper radius, proper volume, and entropy 
               		of the CEH will monotonically increase, asymptoting to a constant. 
			\textit{[See the electronic edition of the journal for a color version of this figure.]}}
               		\label{fig:deht}
       		\end{center}
	\end{figure}
}

The entropy of the CEH is calculated using the Bekenstein-Hawking horizon 
entropy equation as suggested by \citet{Gibbons1977}. 
\begin{eqnarray}
	S_{\mathrm{CEH}} & = & \frac{k c^3}{G \hbar} \frac{A}{4} = \frac{k c^3}{G \hbar} \pi R_{\mathrm{CEH}}^2 \nonumber \\
		& = & 2.6 \pm 0.3 \xt{122}\ k
\end{eqnarray}
Uncertainty in the CEH radius, volume, and entropy are dominated by 
uncertainties in Hubble's constant (Figure \ref{fig:dehh0}).

The CEH monotonically increases, asymptoting to a constant radius and 
entropy slightly larger than its current value (see Figure \ref{fig:deht}). We calculate the
asymptotic radius, volume, and entropy to be
\begin{eqnarray}
	R_{\mathrm{CEH}}(t \rightarrow \infty) & = & 16.4 \pm 0.4\ {Glyr} \nonumber \\   			
			& = & 1.55 \pm 0.04 \xt{26}\ m   			
\end{eqnarray}
\begin{eqnarray}
	V_{\mathrm{CEH}}(t \rightarrow \infty) & = & 1.84 \pm 0.15 \xt{4}\ {Glyr}^3 \nonumber \\ 	
			& = & 1.56 \pm 0.13 \xt{79}\ m^3  		
\end{eqnarray}
\begin{eqnarray}
	S_{\mathrm{CEH}}(t \rightarrow \infty) & = & 2.88 \pm 0.16 \xt{122}\ k. 		
\end{eqnarray}

\bibliographystyle{apj}
\bibliography{entropy}

\begin{thebibliography}{47}
\expandafter\ifx\csname natexlab\endcsname\relax\def\natexlab#1{#1}\fi

\bibitem[{Abe {et~al.}(2008)}]{Abe2008}
Abe, S. {et~al.} 2008, Phys. Rev. Lett., 100, 221803

\bibitem[{{Adams} \& {Laughlin}(1997)}]{Adams1997}
{Adams}, F.~C. \& {Laughlin}, G. 1997, Reviews of Modern Physics, 69, 337

\bibitem[{Adamson {et~al.}(2008)}]{Adamson2008}
Adamson, P. {et~al.} 2008, Phys. Rev. Lett., 101, 131802

\bibitem[{{Balbus} \& {Hawley}(2002)}]{Balbus2002}
{Balbus}, S.~A. \& {Hawley}, J.~F. 2002, \apj, 573, 749

\bibitem[{{Basu} \& {Lynden-Bell}(1990)}]{Basu1990}
{Basu}, B. \& {Lynden-Bell}, D. 1990, \qjras, 31, 359

\bibitem[{{Bekenstein}(1973)}]{Bekenstein1973}
{Bekenstein}, J.~D. 1973, \prd, 7, 2333

\bibitem[{{Bekenstein}(1974)}]{Bekenstein1974}
---. 1974, \prd, 9, 3292

\bibitem[{{Binney} \& {Tremaine}(2008)}]{Binney2008}
{Binney}, J. \& {Tremaine}, S. 2008, {Galactic Dynamics: Second Edition}
  (Princeton University Press)

\bibitem[{{Bousso} {et~al.}(2007){Bousso}, {Harnik}, {Kribs}, \&
  {Perez}}]{Bousso2007}
{Bousso}, R., {Harnik}, R., {Kribs}, G.~D., \& {Perez}, G. 2007, \prd, 76,
  043513

\bibitem[{{Cleveland} {et~al.}(1998){Cleveland}, {Daily}, {Davis}, {Distel},
  {Lande}, {Lee}, {Wildenhain}, \& {Ullman}}]{Cleveland1998}
{Cleveland}, B.~T., {Daily}, T., {Davis}, R.~J., {Distel}, J.~R., {Lande}, K.,
  {Lee}, C.~K., {Wildenhain}, P.~S., \& {Ullman}, J. 1998, \apj, 496, 505

\bibitem[{{Coleman} \& {Roos}(2003)}]{Coleman2003}
{Coleman}, T.~S. \& {Roos}, M. 2003, \prd, 68, 027702

\bibitem[{{Davis} {et~al.}(2003){Davis}, {Davies}, \& {Lineweaver}}]{Davis2003}
{Davis}, T.~M., {Davies}, P.~C.~W., \& {Lineweaver}, C.~H. 2003, Classical and
  Quantum Gravity, 20, 2753

\bibitem[{{Egan} \& {Lineweaver}(2010)}]{EganLineweaver2009b}
{Egan}, C.~A. \& {Lineweaver}, C.~H. 2010, in preparation

\bibitem[{{Elmegreen}(2007)}]{Elmegreen2007}
{Elmegreen}, B.~G. 2007, in Astronomical Society of the Pacific Conference
  Series, Vol. 362, The Seventh Pacific Rim Conference on Stellar Astrophysics,
  ed. Y.~W. {Kang}, H.-W. {Lee}, K.-C. {Leung}, \& K.-S. {Cheng}, 269--+

\bibitem[{{Frampton}(2009{\natexlab{a}})}]{Frampton2009}
{Frampton}, P.~H. 2009{\natexlab{a}}, ArXiv, 0904.2934

\bibitem[{{Frampton}(2009{\natexlab{b}})}]{Frampton2009b}
---. 2009{\natexlab{b}}, Journal of Cosmology and Astro-Particle Physics, 10,
  16

\bibitem[{{Frampton} {et~al.}(2009){Frampton}, {Hsu}, {Kephart}, \&
  {Reeb}}]{Frampton2008}
{Frampton}, P.~H., {Hsu}, S.~D.~H., {Kephart}, T.~W., \& {Reeb}, D. 2009,
  Classical and Quantum Gravity, 26, 145005

\bibitem[{{Frampton} \& {Kephart}(2008)}]{Frampton2008b}
{Frampton}, P.~H. \& {Kephart}, T.~W. 2008, Journal of Cosmology and
  Astro-Particle Physics, 6, 8

\bibitem[{{Frautschi}(1982)}]{Frautschi1982}
{Frautschi}, S. 1982, Science, 217, 593

\bibitem[{{Fryer} \& {Kalogera}(2001)}]{Fryer2001}
{Fryer}, C.~L. \& {Kalogera}, V. 2001, \apj, 554, 548

\bibitem[{{Fukugita} \& {Peebles}(2004)}]{Fukugita2004}
{Fukugita}, M. \& {Peebles}, P.~J.~E. 2004, \apj, 616, 643

\bibitem[{{Gibbons} \& {Hawking}(1977)}]{Gibbons1977}
{Gibbons}, G.~W. \& {Hawking}, S.~W. 1977, \prd, 15, 2738

\bibitem[{{Gnedin} \& {Gnedin}(1998)}]{Gnedin1998}
{Gnedin}, N.~Y. \& {Gnedin}, O.~Y. 1998, \apj, 509, 11

\bibitem[{{Graham} {et~al.}(2007){Graham}, {Driver}, {Allen}, \&
  {Liske}}]{Graham2007}
{Graham}, A.~W., {Driver}, S.~P., {Allen}, P.~D., \& {Liske}, J. 2007, \mnras,
  378, 198

\bibitem[{{Guth}(1981)}]{Guth1981}
{Guth}, A.~H. 1981, \prd, 23, 347

\bibitem[{{Hawking}(1976)}]{Hawking1976}
{Hawking}, S.~W. 1976, \prd, 13, 191

\bibitem[{{Heger} {et~al.}(2005){Heger}, {Woosley}, \& {Baraffe}}]{Heger2005}
{Heger}, A., {Woosley}, S.~E., \& {Baraffe}, I. 2005, in Astronomical Society
  of the Pacific Conference Series, Vol. 332, The Fate of the Most Massive
  Stars, ed. R.~{Humphreys} \& K.~{Stanek}, 339--+

\bibitem[{{Hulse} \& {Taylor}(1975)}]{HulseTaylor1975}
{Hulse}, R.~A. \& {Taylor}, J.~H. 1975, \apjl, 195, L51

\bibitem[{{Kolb} \& {Turner}(1981)}]{Kolb1981}
{Kolb}, E.~W. \& {Turner}, M.~S. 1981, \nat, 294, 521

\bibitem[{{Kolb} \& {Turner}(1990)}]{Kolb1990}
---. 1990, {The early universe} (Frontiers in Physics, Reading, MA:
  Addison-Wesley, 1988, 1990)

\bibitem[{{Linde}(1982)}]{Linde1982}
{Linde}, A.~D. 1982, Physics Letters B, 108, 389

\bibitem[{{Linde}(2009)}]{Linde2009}
---. 2009, private communication

\bibitem[{{Lineweaver} \& {Egan}(2008)}]{LineweaverEgan2008}
{Lineweaver}, C.~H. \& {Egan}, C.~A. 2008, Physics of Life Reviews, 5, 225

\bibitem[{{Lynden-Bell}(1967)}]{LyndenBell1967}
{Lynden-Bell}, D. 1967, \mnras, 136, 101

\bibitem[{{Mather} {et~al.}(1994){Mather}, {Cheng}, {Cottingham}, {Eplee},
  {Fixsen}, {Hewagama}, {Isaacman}, {Jensen}, {Meyer}, {Noerdlinger}, {Read},
  {Rosen}, {Shafer}, {Wright}, {Bennett}, {Boggess}, {Hauser}, {Kelsall},
  {Moseley}, {Silverberg}, {Smoot}, {Weiss}, \& {Wilkinson}}]{Mather1994}
{Mather}, J.~C., {Cheng}, E.~S., {Cottingham}, D.~A., {Eplee}, Jr., R.~E.,
  {Fixsen}, D.~J., {Hewagama}, T., {Isaacman}, R.~B., {Jensen}, K.~A., {Meyer},
  S.~S., {Noerdlinger}, P.~D., {Read}, S.~M., {Rosen}, L.~P., {Shafer}, R.~A.,
  {Wright}, E.~L., {Bennett}, C.~L., {Boggess}, N.~W., {Hauser}, M.~G.,
  {Kelsall}, T., {Moseley}, Jr., S.~H., {Silverberg}, R.~F., {Smoot}, G.~F.,
  {Weiss}, R., \& {Wilkinson}, D.~T. 1994, \apj, 420, 439

\bibitem[{{Mather} {et~al.}(1999){Mather}, {Fixsen}, {Shafer}, {Mosier}, \&
  {Wilkinson}}]{Mather1999}
{Mather}, J.~C., {Fixsen}, D.~J., {Shafer}, R.~A., {Mosier}, C., \&
  {Wilkinson}, D.~T. 1999, \apj, 512, 511

\bibitem[{{Peacock}(1999)}]{Peacock1999}
{Peacock}, J.~A. 1999, {Cosmological Physics} (Cambridge University Press)

\bibitem[{{Penrose}(1979)}]{Penrose1979}
{Penrose}, R. 1979, in General Relativity: An Einstein centenary survey, ed.
  S.~W. {Hawking} \& W.~{Israel}, 581--638

\bibitem[{{Penrose}(1987)}]{Penrose1987}
{Penrose}, R. 1987, in Three hundred years of gravitation, 17--49

\bibitem[{{Penrose}(2004)}]{Penrose2004}
{Penrose}, R. 2004, {The road to reality: a complete guide to the laws of the
  universe} (London: Jonathan Cape)

\bibitem[{{Seljak} {et~al.}(2006){Seljak}, {Slosar}, \&
  {McDonald}}]{Seljak2006}
{Seljak}, U., {Slosar}, A., \& {McDonald}, P. 2006, Journal of Cosmology and
  Astro-Particle Physics, 10, 14, all data - LyA

\bibitem[{{Spergel} {et~al.}(2007){Spergel}, {Bean}, {Dor{\'e}}, {Nolta},
  {Bennett}, {Dunkley}, {Hinshaw}, {Jarosik}, {Komatsu}, {Page}, {Peiris},
  {Verde}, {Halpern}, {Hill}, {Kogut}, {Limon}, {Meyer}, {Odegard}, {Tucker},
  {Weiland}, {Wollack}, \& {Wright}}]{Spergel2007}
{Spergel}, D.~N., {Bean}, R., {Dor{\'e}}, O., {Nolta}, M.~R., {Bennett}, C.~L.,
  {Dunkley}, J., {Hinshaw}, G., {Jarosik}, N., {Komatsu}, E., {Page}, L.,
  {Peiris}, H.~V., {Verde}, L., {Halpern}, M., {Hill}, R.~S., {Kogut}, A.,
  {Limon}, M., {Meyer}, S.~S., {Odegard}, N., {Tucker}, G.~S., {Weiland},
  J.~L., {Wollack}, E., \& {Wright}, E.~L. 2007, \apjs, 170, 377

\bibitem[{{Steinhardt}(2009)}]{Steinhardt2009}
{Steinhardt}, P. 2009, private communication

\bibitem[{{Strominger} \& {Vafa}(1996)}]{Strominger1996}
{Strominger}, A. \& {Vafa}, C. 1996, Physics Letters B, 379, 99

\bibitem[{{Susskind}(1995)}]{Susskind1995}
{Susskind}, L. 1995, Journal of Mathematical Physics, 36, 6377

\bibitem[{{'t Hooft}(1993)}]{tHooft1993}
{'t Hooft}, G. 1993, ArXiv gr-qc/9310026

\bibitem[{{Weisberg} \& {Taylor}(2005)}]{Weisberg2005}
{Weisberg}, J.~M. \& {Taylor}, J.~H. 2005, in Astronomical Society of the
  Pacific Conference Series, Vol. 328, Binary Radio Pulsars, ed. F.~A. {Rasio}
  \& I.~H. {Stairs}, 25--+

\end{thebibliography}

\end{document}